\documentclass[letterpape,11pt]{article}
\pdfoutput = 1 
    
\usepackage{physics}    
\usepackage{jheppub} 
\usepackage[utf8]{inputenc}
\usepackage{amsmath}
\usepackage{slashed}
\newcommand{\nl}{\nonumber \\[5pt]}

\newcommand{\bmat}[1]{\boldsymbol{#1}}

\newcommand{\lb}{\Big{\lbrack}}
\newcommand{\rb}{\Big{\rbrack}}
\newcommand{\lp}{\Big{(}}
\newcommand{\rp}{\Big{)}}
\newcommand{\lbc}{\Big{\lbrace}}
\newcommand{\rbc}{\Big{\rbrace}}
\newcommand{\Bvert}{\Big{\vert}}

\title{TMD factorization for dijet and heavy-meson pair in DIS}

\author[a]{Rafael F. del Castillo,}
\author[b]{Miguel G. Echevarria,}
\author[c]{Yiannis Makris}
\author[a]{and Ignazio Scimemi}
\affiliation[a]{Dpto. de F\'{i}sica Te\'{o}rica \& IPARCOS, Universidad Complutense de Madrid, E-28040 Madrid, Spain}
\affiliation[b]{Dpto. de F\'{i}sica y Matem\'aticas, Universidad de Alcal\'{a}, 28805 Alcal\'{a} de Henares (Madrid), Spain}
\affiliation[c]{INFN Sezione di Pavia, via Bassi 6, I-27100 Pavia, Italy}

\emailAdd{raffer06@ucm.es}
\emailAdd{m.garciae@uah.es}
\emailAdd{yiannis.makris@pv.infn.it}
\emailAdd{ignazios@ucm.es}

\abstract{We study a transverse momentum dependent (TMD) factorization framework for the processes of dijet and heavy-meson pair production in deep-inelastic-scattering in an electron-proton collider, considering the measurement of the transverse momentum imbalance of the two hard probes in the Breit frame. 
For the factorization theorem we employ soft-collinear and boosted-heavy-quark effective field theories. 
The factorized cross-section for both processes is sensitive to gluon unpolarized and linearly polarized TMD distributions and requires the introduction of a new soft function. 
We calculate the new soft function here at one-loop, regulating rapidity divergences with the $\delta$-regulator. 
In addition, using a factorization consistency relation and a universality argument regarding the heavy-quark jet function, we obtain the anomalous dimension of the new soft function at two and three loops.

}

\date{\today}

\begin{document}

\maketitle
%%%%%%%%%%%%%%%%%%%%%%%%%%%%%%%%%%%%%%%%%%%%%%%%%%%%%%%%%%%%%%%%%%%%%%%%%%%%%%%%%%
\section{Introduction}
It is well known that gluons are an essential constituent of nuclei and that the gluon  parton distribution functions (PDFs) can numerically be much bigger than the corresponding quark  distributions, especially  when the parton energy fraction is  small. 
Gluon transverse momentum dependent distributions (TMDs) are also expected to be similarly enhanced, however they result to be difficult to access due to the lack of clean processes where the factorization of the cross-section holds and incoming gluons constitute the dominant effect. 
An example of such a process is the Higgs production in hadronic colliders~\cite{Gao:2005iu,Chiu:2012ir, Echevarria:2015uaa,Neill:2015roa,Gutierrez-Reyes:2019rug}.  
However, extractions of gluon TMDs from the Higgs transverse-momentum spectrum is challenging due to the nature of the scalar boson and its large mass  (f.i. see also \cite{Monni:2019yyr} including jet veto considerations and \cite{Chen:2018pzu} which do not include all TMD effects).
Even for the relatively clean process of Higgs spectrum, both the unpolarized and linearly polarized gluon distributions appear in the leading power factorization of the cross-section in the $q_T/M_H$ expansion, where $q_T$ and $M_H$ are the boson transverse momentum and mass, respectively.
The absence of a color neutral scalar at low energies has driven the attention to quarkonium production both in semi-inclusive deep inelastic scattering (SIDIS) at Electron Ion Collider (EIC) and LHC
~\cite{Echevarria:2015uaa,Mulders:2000sh,Boer:2012bt,Ma:2012hh,Zhang:2014vmh,Ma:2015vpt, Boer:2015uqa,Bain:2016rrv,Mukherjee:2015smo,Mukherjee:2016cjw,Lansberg:2017tlc,Lansberg:2017dzg,Bacchetta:2018ivt,Hadjidakis:2018ifr,DAlesio:2019qpk,Echevarria:2019ynx,Fleming:2019pzj,Scarpa:2019fol,Grewal:2020hoc,Boer:2020bbd,Echevarria:2020qjk}. 
However, the factorization of these processes  is also challenging (see \cite{Echevarria:2019ynx,Fleming:2019pzj}) and a series of QCD effects are present because of the color structure of quarkonia and the complexity of the non-relativistic expansion (commonly used in quarkonium production studies).

 \begin{figure}
\begin{center}
\includegraphics[width=\textwidth]{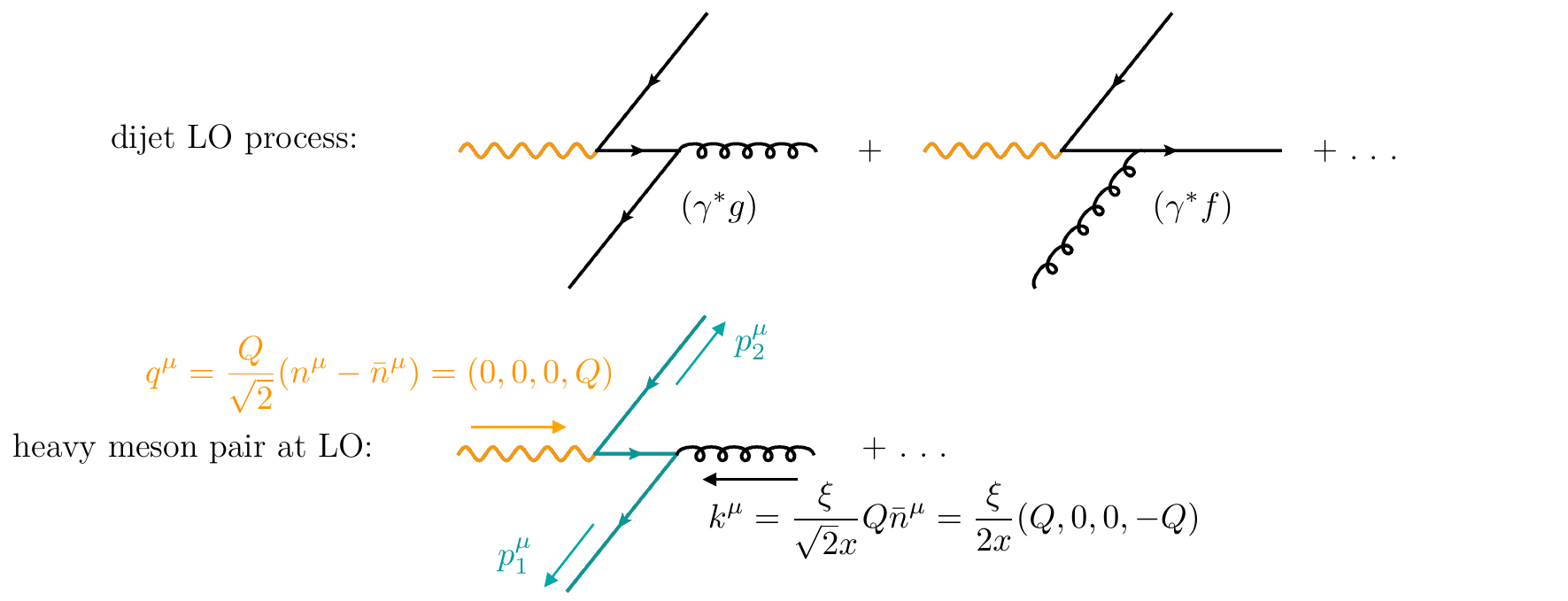}
\caption{\label{fig:LO} Example LO diagrams for the two processes. The momenta $q^{\mu}$ and $k^{\mu}$ (corresponding to the photon and incoming parton momenta respectively) are expressed in the Breit frame.}
\end{center}
\end{figure}

In this work we consider two alternative processes which are presently attracting increasing attention: the dijet~\cite{Dominguez:2010xd} and heavy-meson pair~\cite{Zhu:2013yxa, Zhang:2017uiz, Boer:2010zf} production in an electron-hadron collider, as generated by the $(\gamma^* g)$ and/or $(\gamma^* f)$ hard interactions. 
The processes are
\begin{align}\label{eq:1}
\ell + h&\to  \ell'+J_1 +J_2+X,\;&\text{and}& \; & \ell + h&\to  \ell'+ H + \bar{H}+X
\,,
\end{align}
where $\ell$ and $\ell'$ are the initial and final state leptons, $h$ is the colliding hadron, and $J_i$ and $H/\bar{H}$ are the jets and heavy mesons, respectively. 
All undetected particles in \eqref{eq:1} are represented by $X$.

Dijet production has been the object of several studies at the kinematics of the future
electron ion collider (EIC), as it is sensitive to polarized and unpolarized gluon TMDs~\cite{Chu:2017mnm,Dumitru:2018kuw,Zheng:2018ssm,Page:2019gbf}. 
The produced jets analyzed in the Breit frame have typically a $p_T\in [5,\, 40]$~GeV and are found in the central rapidity region. 
Recent studies (see for example \cite{Page:2019gbf}) suggest that the experimental observation of the dijet imbalance is possible at the future EIC. 
The kinematic constraints we consider here for the dijet process need to be such that do not create hierarchies among the partonic Mandelstam variables, i.e. we demand that $\hat{s} \sim |\hat{t}| \sim |\hat{u}|$. 
If such hierarchies exist, they will induce large logarithms in the hard factor of the cross-section and can potentially ruin the convergence of perturbative expansion, unless further resummation/refactorization of the hard factor is performed.

The heavy-meson pair case is instead experimentally more challenging due to the necessity to reconstruct the momenta of the heavy meson from its decay products. 
In addition, the large energy required to produce a boosted heavy-meson pair makes the process less likely to be observed compared to the dijet production process. 
On the other hand, recent investigations using monte-carlo generators suggest that for charmed mesons this observable could be possible.  From experimental perspective the charm reconstruction have been investigated in refs.~\cite{Arratia:2020azl, Chudakov:2016ytj}.  In ref.~\cite{Li:2020zbk} the charm production rates have been investigated at the LO and NLO QCD for $ep \to c/\bar{c} + X $.
The factorization we construct below requires the transverse momenta of the heavy mesons, $p_T^{H/\bar{H}}$, be parametrically larger than their mass, $m_H$, i.e. $p^{H/\bar{H}}_T \gg m_H$, although an alternative factorization can be constructed when this condition is violated. The details of such factorization involve a hard  function which includes all heavy-quark mass dependence. It also requires a different soft function for which the directions of the quark-antiquark pair are not light-like since the heavy mesons are not boosted to the massless limit. We will not pursue this factorization here, but for a relevant study see ref.~\cite{Zhu:2013yxa}.

At leading order (LO), and ignoring the intrinsic momentum of partons inside the target hadron, the two hard-scattering processes are schematically shown in fig.~\ref{fig:LO}.  
In the case of jets we have that the initial parton can be either a gluon or a quark, while in the heavy-meson case only the gluon initial state is relevant.\footnote{In principle one may consider the case of incoming quark and outgoing gluon which then fragments into a heavy meson. However, in order to access the TMD region (small $\bmat{r}_T$) this fragmentation needs to occurs near threshold, as we discuss later in the main sections, and gluon fragmentation in the kinematic end-point is power suppressed as is discussed both theoretically and phenomenologically in refs.\cite{Fickinger:2016rfd, Anderle:2017cgl}.  } 
We consider the differential cross-section
\begin{align}  
\frac{d\sigma
}{dx d\eta_1 d\eta_2 dp_T d\bmat{r}_T} 
\,,
\label{eq:2} 
\end{align}
where $x$ is the Bjorken variable, and $\eta_i$, $ \bmat{r}_T$  and $p_T$ are  respectively the rapidity, the sum of the transverse momenta (with respect to the beam axis) and the average scalar transverse momenta of the two final jets. 
In the Breit frame, where the virtual photon and target-hadron directions are back-to-back, the factorization holds when $| \bmat{r}_T| \ll p_T$.  
The factorization of the cross-section involves the standard TMDPDFs (we have unpolarized and linearly polarized gluon TMDs, and/or quark TMDs), jet or heavy meson distributions, and a new TMD soft function built with Wilson lines aligned along the directions of the incoming hadron and the two outgoing jets. 
This three-direction soft function has some similarities with the one found in vector boson + jet processes in hadronic colliders~\cite{Buffing:2018ggv,Chien:2019gyf}, however the structure of rapidity divergences is very different compared to the soft function discussed in those studies. 
The perturbative calculation of this new TMD soft function is performed here at one-loop using the modified $\delta$-regulator introduced in \cite{Echevarria:2015byo,Echevarria:2016scs}, and we explicitly check the consistency of the factorization at the same order. 
For the photon-gluon-fusion channel, higher orders of the anomalous dimension of the new soft function can be deduced from the consistency of the anomalous dimensions of the factorized cross-section, since for the case of heavy-meson pair production all other pieces of the cross-section are also known at higher orders.

The presence of the new TMD soft function raises the question of the universality of TMDs beyond the conventional processes of Drell-Yan, semi-inclusive DIS (SIDIS) and di-hadron production in electron-positron annihilation~\footnote{In the case of quark TMDs the conventional universality class has been recently expanded to include also semi-inclusive jet production in the Breit frame  and jet-jet or hadron-jet decorrelation in lepton colliders \cite{Gutierrez-Reyes:2018qez, Gutierrez-Reyes:2019vbx, Gutierrez-Reyes:2019msa}.}. 
While the non-perturbative evolution is universal (in the processes we are considering here and other processes such as SIDIS), the non-perturbative corrections to the soft matrix element are not yet connected to other processes. It is therefore non-trivial to independently extract the gluon TMDPDFs. 
However, the size of universality breaking effects can be estimated phenomenologically by comparison with simpler processes~\footnote{For example one may compare lepton-jet decorrelations in the laboratory frame against jet TMDs in the Breit frame.}. 
For a quantitative analysis of these effects further theoretical advancements are needed.

The paper is organized as follows: 
in sec.~\ref{sec:2} we set the notation to be used in the rest of the paper and we give the dijet process factorization theorem.  
In sec.~\ref{sec:3} we extend the discussion to the heavy-meson pair production and we comment on the universality of the heavy-quark jet functions that appear in our factorization theorem, and the corresponding fragmentation shape functions used to describe heavy-meson fragmentation at threshold. 
Finally we conclude in sec.~\ref{sec:4}. 
In appendices~\ref{sec:app-1} and~\ref{sec:app-2} we collected known results from the literature which we use. 
In appendix~\ref{sec:app-3} we give a pedagogical review of some loop calculations made in this work.

%%%%%%%%%%%%%%%%%%%%%%%%%%%%%%%%%%%%%%%%%%%%%%%%%%%%%%%%%%%%%%%%%%%%%%%%%%%%%%%%%%
\section{Dijet imbalance }
\label{sec:2}

In this section we discuss the factorization of the cross-section for the dijet case in DIS within the soft-collinear effective theory (SCET). 
We do not give a detailed derivation of the factorization theorem, but we rather summarize the final result. 
We also present the NLO calculation for the new three-direction soft function and perform a consistency check of our results using the invariance of the cross-section under renormalization group evolution. 
The notation and kinematics that we develop here are also useful for the heavy-meson pair production presented in the subsequent section.
%%%%%%%%%%%%%%%%%%%%%%%%%%%%%%%%%%%%%%%%%%%%%%%%%%%%%%%%%%%%%%%%%%%%%%%%%%%%%%%%%%
\subsection{Notation and kinematics}
 Assuming that the direction of the beam is along  the $\hat{z}$ axis it is useful to define the four-vector \begin{equation}
 n^{\mu} = \frac{1}{\sqrt{2}}(1,0,0,1)\;.
 \end{equation}
 We also define a conjugate vector $\bar{n} ^{\mu}$ by reversing the sign of the spacial coordinates. Thus, $n^{\mu}$ and $\bar{n}^{\mu}$ satisfy,  
\begin{align}
    n^2 &= \bar{n}^2 = 0,\; \quad \bar{n}\cdot n = 1.
\end{align}
Using the vectors $n^{\mu}$ and $\bar{n}^{\mu}$ we can decompose any other four-vector, $p^{\mu}$, into its light-cone components,  
 \begin{align}
    p^\mu &= p_+ \bar{n}^{\mu} + p_- n^{\mu} + p_{\perp}^{\mu} = (p_+,p_-,p_\perp)_{n},
\end{align}
with
\begin{align}
 p_+ = n\cdot p,\;\quad  p_- = \bar{n}\cdot p,\;\quad    p^2 = 2p_+ p_- + p^2_\perp = 2p_+ p_- - \bmat{p}^2.
\end{align}
where we use the notation $\bmat{p} \equiv \vec{p}_{\perp}$. For the direction of the two jets we use $v_1$ and $v_2$, normalized as 
\begin{align}
v_J^2 = \bar{ v}_J^2 = 0, \;\quad v_J\cdot \bar{v}_J  = 1,\;\text{ with }  J =1,2
\,, 
\end{align}
where the conjugate vectors $\bar{v}_J$, as above, are defined by reversing the sign of the spacial components. 
We define the standard Lorentz-invariants,
\begin{align}
    Q^2 = -q^2,\; \quad x = \frac{Q^2}{2 P\cdot q},
\end{align}
where $q^{\mu}$ is the momentum of the virtual photon, $P^{\mu}$ is the momentum of the target hadron.  
In the Breit frame we have $q^{\mu}=(0,0,0,Q)$ and neglecting  mass corrections  we can solve for target hadron  momentum, 
\begin{equation}
P^{\mu} = \frac{1}{2x} (Q,0,0,-Q)\;.
\end{equation}
The ratio of the longitudinal momenta of the incoming parton and the target hadron we denote with $\xi$,
\begin{equation}
  \xi = \frac{k^+}{P^+}\;.
\end{equation}
where $k^{\mu}$ the momenta of the parton incoming to the hard process. We can then express the variables $Q$ and $\xi$ in terms of the  Born level kinematics using the pseudo-rapidities, $\eta_1$ and $\eta_2$, and the transverse momentum, $p_T$, of the two outgoing partons,
\begin{align}
    Q &= 2p_T \cosh(\eta_-) \exp(\eta_+),\;& \xi = 2 x \cosh(\eta_+) \exp(-\eta_+)\;,
\end{align}
where 
\begin{equation}
    \eta_{\pm} = \frac{\eta_1 \pm \eta_2}{2}\;,
\end{equation}
In this expressions we have neglected corrections from the target hadron mass. The partonic Mandelstam variables in terms of the same quantities are, 
\begin{align}
    \hat{s} &=(q+k)^2 =  + 4 p_T^2 \cosh^2(\eta_-) \;, \nl
    \hat{t} &=(q-p_2)^2= -4 p_T^2 \cosh(\eta_-) \cosh(\eta_+) \exp(\eta_1)\;, \nl
    \hat{u} &=(q-p_1)^2= -4 p_T^2 \cosh(\eta_-) \cosh(\eta_+) \exp(\eta_2)\;,
\end{align}
where $p_1^{\mu}$ and $p_2^{\mu}$ are the momenta of the outgoing partons. It is easy to check that the  partonic Mandelstam variables satisfy, 
\begin{equation}
    \hat{s} +\hat{t}+\hat{u} = - Q^2\;.
\end{equation}

We denote the transverse momentum imbalance of the two jets  with  $\bmat{r}_T$, where the hard transverse  momentum $p_T$ corresponds,  up-to power corrections, to the average transverse momenta of the two jets,  
\begin{align}
    \bmat{r}_T & = \bmat{p}_{1T} + \bmat{p}_{2T} ,\; &p_T  &= \frac{|\bmat{p}_{1T} | + |\bmat{p}_{2T}| }{2}
\end{align}
where the sub-index 1,2 refers to the final jets. 
At Born level $\bmat{p}_{1T} =  - \bmat{p}_{2T}$ and thus $ \bmat{r}_T  =0 $.  
However, hadronization of the outgoing partons will form jet-like configurations along similar directions and wide angle radiation could escape the jet clustering algorithm, which will then contribute to the imbalance.

\begin{figure}
\begin{center}
\includegraphics[width= 0.58 \textwidth]{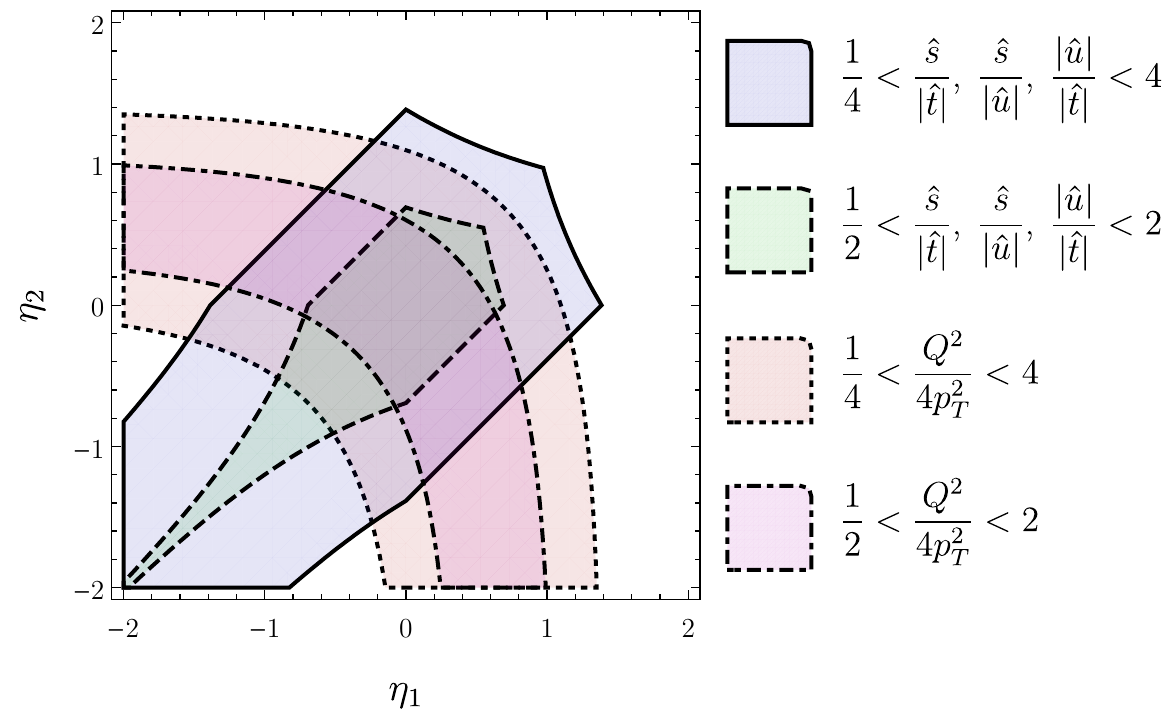}
\caption{\label{fig:rapidity} The jet-rapidity regions for which $\hat{s} \sim |\hat{t}| \sim |\hat{u}|$: blue-solid ($j=4$) and green-dashed ($j=2$) and $Q \sim p_T$: red-dotted $k=4$ and magenta-dotdashed ($k=2$). The overlapping region which is found at central rapidities is where we anticipated small contributions to the cross section beyond factorization.}
\end{center}
\end{figure}

We now consider the kinematic region in which the factorization theorem holds in comparison to the coverage of EIC. We first evaluate the constraints on the rapidities of the two jets, $\eta_1$ and $\eta_2$. To do so we require that $\hat{s} \sim |\hat{t}| \sim |\hat{u}|$ and quantitatively we implement that by imposing,
\begin{equation}\label{eq:stu-consrtaint}
    \frac{1}{j} < \frac{\hat{s}}{|\hat{t}|},\; \frac{\hat{s}}{|\hat{u}|},\; \frac{|\hat{u}|}{|\hat{t}|} < j
\end{equation}
  This constrains the values of rapidities within the blue region as illustrated in figure~\ref{fig:rapidity} for the cases $j=2$ (green-dashed) and $j = 4$ (blue-solid).  In addition to avoid contributions from the resolved photon processes we require $Q \sim p_T$ which quantitatively we implement by imposing,
\begin{equation}\label{eq:Q-consrtaint}
    \frac{1}{k} < \frac{Q^2}{4 p_T^2} < k
\end{equation}
The relevant region in the two jet rapidities is shown as the red-dotted and magenta-dotdashed shaded areas for $k=4$ and $k =2 $ respectively in figure~\ref{fig:rapidity}. The overlap of the regions constraint by \eqref{eq:Q-consrtaint}  and \eqref{eq:stu-consrtaint}  gives the allowed values of rapidites for which the factorization theorem holds and contamination from resolved photo-production processes is minimal. This suggests that the two processes we are considering here, are described by two clearly separated jets (or heavy mesons) within the central rapidity region. As expected, when tighten the constraint (by decreasing the values of $j$ and/or $k$) the relevant region shrinks around the central point, $\eta_1 =\eta_2 =0$.

\begin{figure}[h!]
\begin{center}
\includegraphics[width=  \textwidth]{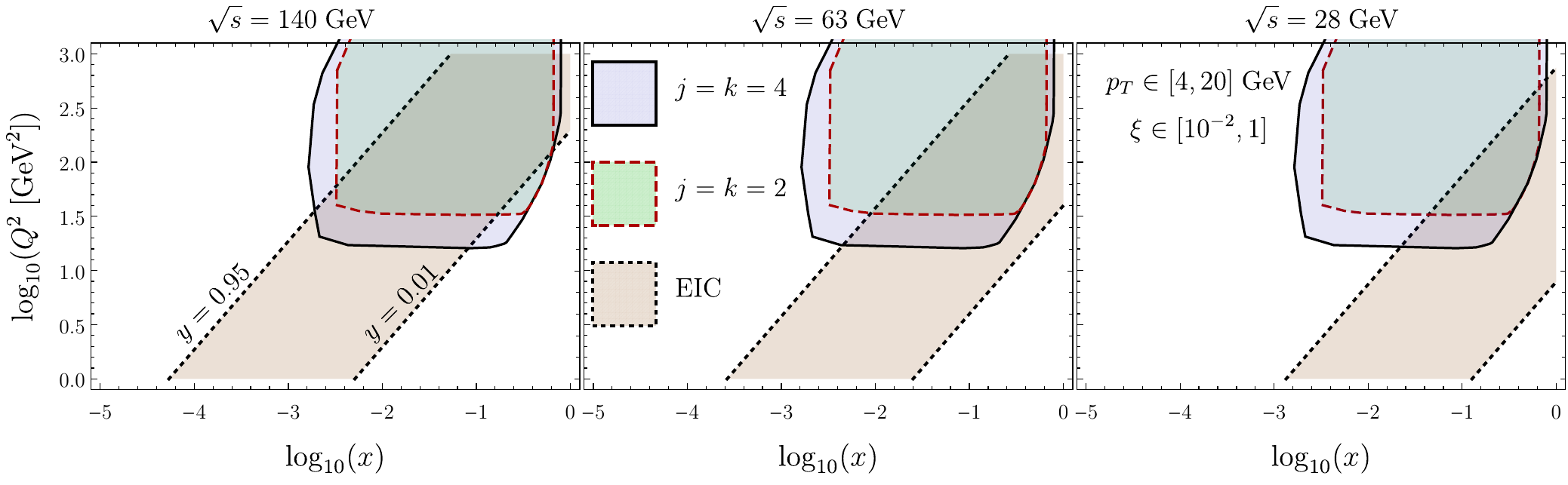}
\caption{\label{fig:coverage} The ($x,Q^2$) coverage of EIC (brown/dashed) compared to the di-jet events with $j = k =4$ factorization regime (blue-solid) and with  $j =k =2$ (green-dashed). The factorization regime has been constructed assuming $p_T \in [4,20]$ GeV, $\xi \in [10^{-2},1]$, and jet rapidities in the overlapping region in figure~\ref{fig:rapidity}.} 
\end{center}
\end{figure}

Constrained within the overlapping region of rapidities  in figure~\ref{fig:rapidity} and for $p_T \in [4,20]$ GeV  and $\xi \in [10^{-2},1]$ we construct the $(x,Q^2)$ values relevant for the process we are considering. We show the corresponding region in ($x,Q^2$) in figure~\ref{fig:coverage} for $j=k=2$ (green-dashed) and  $j=k=4$ (blue-solid).  We also included (brown-shaded area) the expected $(x,Q^2)$  coverage at EIC for three different center of mass energies: $\sqrt{s} = 140$, 63, and 28 GeV. We see that for all three energies there is significant overlap of the investigated process and the EIC coverage, but the overlapping region increases at higher beam energies. At the same time the cases $j=k=2$  and $j=k=4$ does seem to change the overlapping region with the EIC coverage in the small $Q^2$ region. Although the change of the overlapping region is not large, the lost region is statistically important. A further investigation into this using monte carlo event generators will be important  in order to determine the kinematic constraints on the jets, for which a reasonable compromise between statistics and factorization-corrections can be made.\footnote{Note that here for both scenarios we consider $j=k$ even-though $j$ and $k$ are independent variables and maybe chosen to be not equal. They need to be however, order $\mathcal{O}(1)$ numbers. For example, the case $j=2,\;k=4$, which is not shown here, seems to have very little effect on the overlapping region with the EIC coverage, compared to  the case $j=4,\;k=4$.}

%%%%%%%%%%%%%%%%%%%%%%%%%%%%%%%%%%%%%%%%%%%%%%%%%%%%%%%%%%%%%%%%%%%%%%%%%%%%%%%%%%
\subsection{Factorization theorem for dijet production}

There are two channels for the dijet process that we need to consider: a) the gluon-photon fusion channel which corresponds to the partonic process, $\gamma^* g \to f\bar{f}$, and b) the  incoming quark or antiquark channel from the partonic process, $\gamma^* f \to g f$.
 The factorization that we propose  holds when   $| \bmat{r}_T|\ll p_T$ and we treat  the cross-sections only at leading power in an $| \bmat{r}_T|/ p_T$ or $| \bmat{r}_T|/Q$ expansion. Within this constraint, the gluon-photon channel  factorized cross-section is  
\begin{multline}
\label{eq:sgfg}
    \frac{d\sigma (\gamma^* g)}{dx d\eta_1 d\eta_2 dp_T d\bmat{r}_T} =
    \sum_{f} H^{\mu \nu} _{\gamma^*g\to f \bar{f} }(\hat{s},\hat{t}, \hat{u},\mu) \int \frac{d^2 \bmat{b}}{(2\pi)^2}  \, \exp(i \bmat{b} \cdot \bmat{r}_T) \,F_{g, \mu \nu}(\xi, \bmat{b}, \mu,\zeta_1) \\
    \times S_{\gamma g}(\bmat{b},\eta_1, \eta_2,\mu,\zeta_2) \,\lp \mathcal{C}_{f}(\bmat{b},R,\mu)  J_{f}(p_T,R,\mu) \rp \lp \mathcal{C}_{\bar{f}}(\bmat{b},R,\mu)  J_{\bar{f}}(p_T,R,\mu) \rp \;,
\end{multline}
where the sum runs over all light quark and antiquark flavours $f$. 
Here we consider jets with their momentum reconstructed with the so called E-scheme, that is, the momentum of the jets is given by the sum of all jet-constituents. 
For these jets and for small jet radius ($R\ll 1$) the cross-section can be factorized in terms of the collinear-soft function ${\cal C}_i(\bmat{b},R)$, that describes the soft radiation close to the jet boundary and the exclusive jet functions $J_{i}$, that describe the collinear and energetic radiation confined within the jet.\footnote{For recent developments on jet algorithms for DIS see ref.~\cite{Arratia:2020ssx}. We also plan in the near future to complete the NLO calculation of jet functions in central rapidity regions for the Centauro algorithm. } 
These functions are calculated up to NLO for generic $k_T$-type and cone jet algorithms in \cite{Hornig:2016ahz,Buffing:2018ggv}.
The corresponding operator definitions are given in the appendix~\ref{sec:app-1}.
In addition, the factorization theorem contains the dijet soft function $S_{\gamma g}$, which is discussed and calculated at one-loop in next section. 
Finally, we have the gluon TMD function $F_g$, whose operator definition can also be found in appendix~\ref{sec:app-1}.

We notice at this point that the latter two functions, $S_{\gamma g}$ and $F_g$  in \eqref{eq:sgfg}, have an intricate interplay due to the rapidity divergences, which introduces the rapidity scale dependence, $\zeta_{1,2}$, in the corresponding functions. 
In sec.~\ref{sec:zerobin} we explain this issue and the role played by the zero-bin subtractions, which lead to the proper definition of the dijet soft function and the TMDs.

The gluon TMD for an unpolarized proton can be further separated into two pieces, the unpolarized gluon distribution $f_1(\xi,\bmat{b})$  and the linearly polarized gluon contribution $h_{1}^{\perp}(\xi,\bmat{b})$:
\begin{equation}
\label{eq:hadtens}
    F_{g}^{\mu\nu}(\xi,\bmat{b}) = f_1(\xi,\bmat{b})\frac{g_{T}^{\mu\nu}}{d-2}  + h_1^\perp (\xi,\bmat{b})\,\lp \frac{g_{T}^{\mu\nu}}{d-2} + \frac{b^{\mu} b^{\nu}}{\bmat{b}^2}\rp
\,,
\end{equation}
with $g_T^{\mu\nu}=g^{\mu\nu}-n^\mu\bar n^\nu-\bar n^\mu n^\nu$. 
Both of these functions are known perturbatively up to next-to-next-to leading order (NNLO)~\cite{Echevarria:2016scs,Gutierrez-Reyes:2019rug,Luo:2019bmw}. 
The evolution of the TMDs, which is universal (see e.g. \cite{Echevarria:2012pw,Echevarria:2014rua}), is also known up to N$^3$LO~\cite{Echevarria:2015byo,Li:2016ctv,Vladimirov:2016dll}. 
In \eqref{eq:hadtens} we have included only twist-2 TMDs, neglecting higher twists in the TMD expansion. 
This is sufficient since we consider higher-twist functions suppressed, which is consistent with SIDIS studies as in \cite{Scimemi:2019cmh}. 
However, we have no quantitative estimate of these functions and further investigation is important.
The inclusion of the higher-twist contributions is beyond the scope of this work and we leave such considerations for future studies. 
The hard function is then decomposed in two tensor structures: 
\begin{equation}
    H^{\mu\nu}_{\gamma^* g \to f\bar{f}} = \sigma_0^{g U}\, H^U_{\gamma^*g\to f \bar{f} } \frac{g_{T}^{\mu\nu}}{d-2} + \sigma_0^{g L} \,H^L_{\gamma^*g\to f \bar{f} } \lp -\frac{g_{T}^{\mu\nu}}{d-2} +\frac{v^{\mu}_{1T}\, v^{\nu}_{2T} + v^{\mu}_{2T}\, v^{\nu}_{1T} }{2 \;v_{1T} \cdot v_{2T}}\rp 
    \,,
\end{equation}
where we have ignored all terms proportional to the four-vector $n^{\mu}$ (since  they vanish after Lorentz-contraction with the gluon beam function) and any anti-symmetric combinations (since the cross-section is integrated over angles). 
The coefficients $\sigma_0^{g U(L)}$ are introduced such that the leading order hard functions are normalized to the unity, i.e. $H^{U(L)}_{\text{LO}} = 1 +\mathcal{O}(\alpha_s)$. 
With this we can now separate the cross-section into a contribution from the unpolarized gluons and one from the linearly polarized gluons. 
We write the cross-section as
\begin{equation}
    d\sigma (\gamma^* g ) = d\sigma^U (\gamma^* g ) + d\sigma^L (\gamma^* g )
    \,,
\end{equation}
where
\begin{multline}\label{eq:FactGamGU}
    \frac{d\sigma^U  (\gamma^* g )}{dx d\eta_1 d\eta_2 dp_T d\bmat{r}_T} = \sigma_0^{g U}\,\sum_{f} H^{U}_{\gamma^*g\to f \bar{f} }(\hat{s},\hat{t}, \hat{u},\mu) \int \frac{d^2\bmat{b}}{(2\pi)^2}  \, \exp(i \bmat{b} \cdot \bmat{r}_T) \, f_1(\xi, \bmat{b}, \mu,\zeta_1) \\
    \times S_{\gamma g}(\bmat{b},\zeta_2,\mu) \,\lp \mathcal{C}_{f}(\bmat{b},R,\mu)  J_{f}(p_T,R,\mu) \rp \lp \mathcal{C}_{\bar{f}}(\bmat{b},R,\mu)  J_{\bar{f}}(p_T,R,\mu) \rp\;,    
\end{multline}
and
\begin{multline}\label{eq:FactGamGL}
    \frac{d\sigma^L  (\gamma^* g )}{dx d\eta_1 d\eta_2 dp_T d\bmat{r}_T} = \sigma_0^{g L}\,\sum_{f} H^{L}_{\gamma^*g\to f \bar{f} }(\hat{s},\hat{t}, \hat{u},\mu) \int \frac{d^2\bmat{b}}{(2\pi)^2}  \, \exp(i \bmat{b} \cdot \bmat{r}_T) \, h_1^{\perp}(\xi, \bmat{b}, \mu,\zeta_1) \\
    \times \frac{s_{\bmat {b}}^2 - c_{\bmat {b}}^2}{2}\;S_{\gamma g}(\bmat{b},\zeta_2,\mu) \,\lp \mathcal{C}_{f}(\bmat{b},R,\mu)  J_{f}(p_T,R,\mu) \rp \lp \mathcal{C}_{\bar{f}}(\bmat{b},R,\mu)  J_{\bar{f}}(p_T,R,\mu) \rp\;.   
\end{multline}
We used the shorthand notation $s_{\bmat {b}}$ and $c_{\bmat{b} }$ for the sine and cosine of the angle between the vectors $\bmat{b}$ and $\bmat{v}_{1T}$, respectively. 
The hard factors are calculated up to NNLO in the unpolarized case in \cite{Becher:2009th,Becher:2012xr}, while for the linearly polarized gluons they are calculated at LO in \cite{Chien:2020hzh}. 
All hard coefficients are reported in appendix~\ref{sec:app-1}.

Finally, the incoming quark channel has contributions only from the unpolarized gluon jets and the cross-section is given by the following factorized formula:
\begin{multline}\label{eq:FactGamFU}
    \frac{d\sigma^{U} (\gamma^* f)}{dx d\eta_1 d\eta_2 dp_T d\bmat{r}_T} = \sigma_0^{fU}\,\sum_{f} H^{U}_{\gamma^*f\to g f }(\hat{s},\hat{t}, \hat{u},\mu) \int \frac{d^2\bmat{b}}{(2\pi)^2}  \, \exp(i \bmat{b} \cdot \bmat{r}_T) \,F_{f}(\xi, \bmat{b},  \mu,\zeta_1) \\
    \times S_{\gamma f}(\bmat{b},\zeta_2,\mu) \lp \mathcal{C}_{g}(\bmat{b},R,\mu)  J_{g}(p_T,R,\mu) \rp  \lp \mathcal{C}_{f}(\bmat{b},R,\mu)  J_{f}(p_T,R,\mu) \rp,
\end{multline}
where the sum runs over quarks and anti-quarks and $F_f$ is the $f$-flavor quark/antiquark  unpolarized TMDPDF.
%%%%%%%%%%%%%%%%%%%%%%%%%%%%%%%%%%%%%%%%%%%%%%%%%%%%%%%%%%%%%%%%%%%%%%%%%%%%%%%%%%
\subsection{The dijet soft function at NLO}

\begin{figure}
\begin{center}
\includegraphics[width=\textwidth]{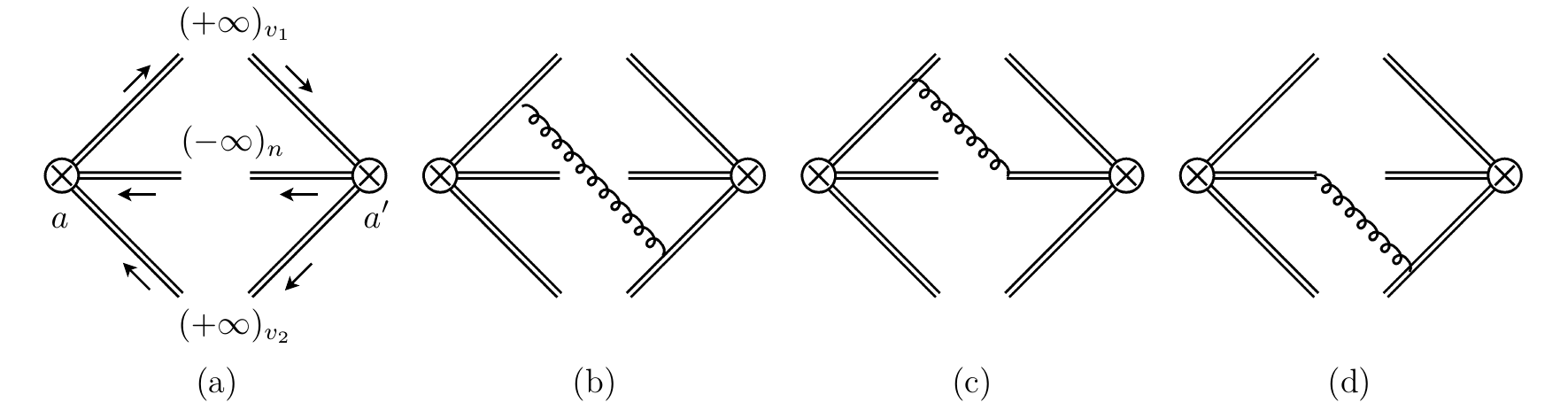}
\caption{\label{fig:soft} 
Tree level soft function is shown in diagram (a). 
Each double line represents a Wilson line whose pointing direction is also reported as $(\pm \infty)_u$, $u=n,\;v,\; \bar v$; 
$a$ and $a'$ are color indices. 
Diagrams (b), (c) and (d) also contribute to the NLO soft function.
Virtual contributions vanish and therefore they are not shown here. 
Mirror diagrams are also not shown.}
\end{center}
\end{figure}

The only new matrix element in the previous section is the soft function. 
Here we give the operator matrix element definition of the soft function and we proceed with the NLO calculation. 
The details of the calculation are collected in the appendix.  
We start defining the soft function for the photon-gluon fusion process:
\begin{multline}
\label{SF}
   \hat S_{\gamma g}(\bmat{b}) = \frac{1}{C_F C_A} \langle 0| \mathcal{S}^\dagger_n(\bmat{b},-\infty)_{ca'}\text{Tr}\lb S_{v_2}(+\infty,\bmat{b})T^{a'}S^\dagger_{v_1} (+\infty,\bmat{b}) \\
    \times S_{v_1}(+\infty,0)T^{a}S^{\dagger}_{v_2} (+\infty,0)\rb \mathcal{S}_n(0,-\infty)_{ac}|0 \rangle.
\end{multline}
The  soft function corresponding to the case of incoming  quark or antiquark  is obtained with the exchange
\begin{equation}
\hat S_{\gamma f} = \hat S_{\gamma g} (n \leftrightarrow v_2)
\,.
\end{equation}
The Wilson lines are defined as
\begin{align}
\label{Wilsonlines}
    S_v(+\infty,\xi) &= P\text{exp}\lb-ig\int_0^{+\infty}d\lambda\; v \cdot A(\lambda v + \xi)\rb \,,\nl
    S^{\dagger}_{\bar{v}}(+\infty,\xi)& = P\text{exp}\lb ig\int_0^{+\infty}d\lambda\; \bar{v} \cdot A(\lambda \bar{v} + \xi)\rb\,, \nl
    S_n(+\infty,\xi) &= \lim_{\delta^+\to 0}P\text{exp}\lb-ig\int_0^{+\infty}d\lambda\; n \cdot A(\lambda n + \xi) e^{-\delta^+ \lambda}\rb \,.
\end{align}
We omit here $T$-Wilson lines which are needed in singular gauges. 
We also distinguish Wilson lines in the adjoint and fundamental SU($N_c$) representations using $\mathcal{S}$ and $S$ respectively.  
Note that the $\delta$-regulator is introduced only in $S_n(\mathcal{S}_n)$. 
We write the soft function as a series in $a_s=\alpha_s/(4\pi)$
 \begin{align}
  \hat S=\sum_{m=0}^\infty a_s^m \hat S^{[m]}  \,. 
 \end{align}
At tree level $\hat S^{[0]}=1$, and at one-loop only the diagrams with a real gluon give a non-zero result. 
We can therefore write the one-loop soft function as a sum of diagrams with one real gluon exchange between the three soft Wilson lines, 
\begin{equation}
    \hat S^{[1]}=\frac{1}{2} \sum_{ \substack {i \ne j \\ j \in \{ 1, 2, B \} } }  C^{ij}\,\hat S^{[1]}_{ij} \;,
\end{equation} 
where the suffix indicates the  Wilson lines  connected by the exchanged gluon (1 and 2 for the two jets and $B$ for the beam), including the mirror diagrams as well (note that $C^{ij} = C^{ji}$ and $ \hat S^{[1]}_{ij} = \hat S^{[1]}_{ji}$). 
Notice that the mirror diagrams for $\hat{S}_{JB}$ will introduce an additional  $i \pi$ component, as discussed in  more detail in the appendix. The coefficients $C^{ij}$ are the color factors and they are different for the two channels $\gamma^*g$ and $\gamma^* f$,
\begin{align}
C^{1B}_{\gamma g} &= C^{2B}_{\gamma g} =  C_A\;, & C^{12}_{\gamma g} &= 2C_F - C_A, \nl
C^{2B}_{\gamma f} &=   C^{12}_{\gamma f} =  C_A\;, & C^{1B}_{\gamma f} &=  2C_F - C_A.
\end{align}
The relevant diagrams are shown in fig.~\ref{fig:soft} and the corresponding contributions are 
 \begin{align}
  	\hat S_{JB}^{[1]} &=  \pi (4\pi)^2 \mu^{2\epsilon} (n\cdot v_J) I_{JB}+\text{ mir. diag.}\;, &
 \hat  S_{12}^{[1]} &=  \pi (4\pi)^2 \mu^{2\epsilon} (v_1 \cdot v_2) I_{12} +\text{ mir. diag.},  
\end{align}
where $J=1,2$ and the corresponding integrals are\begin{align}\label{eq:softintegrals}
 	I_{JB} &= \int\frac{d^d k}{(2\pi)^d}\frac{e^{i \bmat{k}\cdot \bmat{b}}\delta(k^2)\theta(k_+)}{(n\cdot k+i\delta^+)(v_J\cdot k)} \;, & 
    	I_{12} &= \int\frac{d^d k}{(2\pi)^d}\frac{e^{i \bmat{k}\cdot \bmat{b}}\delta(k^2)\theta(k_+)}{(v_1\cdot k)(v_2\cdot k)}
	\,,
 \end{align}
 with $d = 4 - 2\epsilon$.
The results at all orders in $\epsilon$ are
\begin{equation}
  \hat  S_{JB}^{[1]} = - 2 \frac{\mu^{2\epsilon }   B^\epsilon }{(4\pi)^{-\epsilon}} \,\Gamma(-\epsilon) \lb \ln \lp - \frac{i(\bmat{v}_J\cdot \bmat{b})\,\delta^+}{n\cdot v_J}\rp+\gamma_E \rb  ,
\end{equation}
and
\begin{equation}
 \hat   S_{12}^{[1]} =     2 \frac{\mu^{2\epsilon }   B^\epsilon }{(4\pi)^{-\epsilon}} \,\Gamma(-\epsilon) \lb \Gamma(-\epsilon) \Gamma(1+\epsilon) \lp \frac{1+ A_{\bmat{b}}}{-A_{\bmat{b}}}\rp ^{\epsilon} 
    - \frac{\Gamma(-1-\epsilon)}{\Gamma(-\epsilon)}A_{\bmat{b}}\,_{2} F_{1}(1,1,2+\epsilon,-A_{\bmat{b}} ) \rb,
\end{equation}
 with the  shorthand notation,
\begin{align}
\label{eq:notation-AB}
A_{{\bmat{b}}}& = \frac{(v_1 \cdot v_2)}{2 \,(v_{1} \cdot \hat{b})   \,(v_{2} \cdot \hat{b})  } = - \frac{\hat{s}}{4 \,p_T^2 \,c^2_b},\;& B& = \frac{| \bmat{b} |^2}{4},
\end{align}
and  $\hat{b} = (0,0,\bmat{b})/|\bmat{b} |$. Note that $A_{\bmat{b}}$ is a function of the angle $\phi_{bJ}$ only and it does not depend on  $|\bmat{b}|$. 
$A_{\bmat{b}}$ is also  dimensionless, longitudinal boost invariant, and it is bounded to negative values less than $-1$, i.e.,  $A_{\bmat{b}}  \leq -1$. The function $_2 F_1$ is the standard hypergeometric function and the $\epsilon$ expansion for  this function can be written as follows
\begin{equation}
\,_2 F_1(1,1,2+\epsilon, - z) =  \frac{1}{z} \ln (1+ z) + \frac{\epsilon}{z} \lb  -\frac{\pi^2}{6} + \lp 1+ \ln \lp\frac{1+z}{z} \rp  \rp \ln (1+z) + \text{Li}_2 \lp \frac{1}{1+z}\rp  \rb + \mathcal{O}(\epsilon^2).
\end{equation}
Adding all contributions and expanding in $\epsilon$ we obtain the bare soft function. For the $\gamma^* g$-channel we have:
\begin{multline}
\hat S^{\text{bare}}_{\gamma g} (\bmat{b}) =  \hat S^{\text{finite}}_{\gamma g} (\bmat{b}) + a_s \lbc 
C_A \lb -\frac{2}{\epsilon^2} + \frac{2}{\epsilon}\lp      2\ln \lp  \frac{ \sqrt{2} \, \delta^+}{\mu} \rp  + \ln (2 A_n)   \rp   \rb \\[5pt] 
+ 2 C_F \lb \frac{2}{\epsilon^2} + \frac{2}{\epsilon} \ln \lp \frac{B \,\mu^2 \,e^{2\gamma_E}}{- A_{\bmat{b}}} \rp     \rb
\rbc ,
\end{multline}
 for the $\gamma^* f$-channel: 
 \begin{multline}
 \hat S^{\text{bare}}_{\gamma f} (\bmat{b}) =  \hat S^{\text{finite}}_{\gamma f} (\bmat{b}) + a_s \lbc 
  C_A \lb \frac{2}{\epsilon^2} +\frac{2}{\epsilon} \lp \ln \frac{(n \cdot v_1) (\bmat{v}_2 \cdot \bmat{b})}{(n \cdot v_2) (\bmat{v}_1 \cdot \bmat{b})} +\ln \lp \frac{B \mu^2 e^{2\gamma_E}}{-A_{\bmat{b}}}\rp  \rp  \rb \\[5pt]
  +\frac{4}{\epsilon}C_F \ln \lp -\frac{i\,\bmat{v_1 \cdot \bmat{b}} \,\delta^+ e^{\gamma_E} }{n\cdot v_1} \rp  
 \rbc ,
 \end{multline}
where the finite part of the soft function is
\begin{multline}
\hat S^{\text{finite}}_{\gamma g} (\bmat{b}) = 1 +  a_s\lbc
C_A  \lb \ln (B \,\mu^2 e^{2\gamma_E}) \lp  \ln (B \mu^2 e^{2\gamma_E})  + 4\ln \lp \frac{\sqrt{2}\,\delta^+}{\mu} \rp + 2\ln(2A_n)  \rp  - \ln^2(-A_{\bmat{b}}) \\[5pt] -\frac{\pi^2}{6} - 2  \text{Li}_2(1+A_{\bmat{b}})   \rb 
+ C_F  \lb \frac{\pi^2}{3}  + 2 \ln^2 \lp \frac{B \mu^2 e^{2\gamma_E}}{ -A_{\bmat{b}}}\rp + 4 \text{Li}_2(1+A_{\bmat{b}})  \rb
\rbc ,
\end{multline}
and
\begin{multline}
\hat S^{\text{finite}}_{\gamma f} (\bmat{b}) = 1 + a_s \lbc
C_A \lb  \frac{\pi^2}{6}  +  \ln^2 \lp \frac{B \mu^2 e^{2\gamma_E}}{ -A_{\bmat{b}}}\rp + 2 \text{Li}_2(1+A_{\bmat{b}}) + 2\ln(B \mu^2 e^{2\gamma_E} )   \ln \frac{(n \cdot v_1) (\bmat{v}_2 \cdot \bmat{b})}{(n \cdot v_2) (\bmat{v}_1 \cdot \bmat{b})}   \rb \\[5pt]
+4 C_F \ln(B \mu^2 e^{2\gamma_E} )   \ln \lp - \frac{i\,\bmat{v_1 \cdot \bmat{b}} \,\delta^+ e^{\gamma_E} }{n\cdot v_1} \rp
 \rbc .
\end{multline}
To simplify the results we have used the notation
\begin{equation}
A_n = \frac{ (v_1 \cdot v_2)}{2\, ( v_1 \cdot n ) ( v_2 \cdot n)}.
\end{equation}

%%%%%%%%%%%%%%%%%%%%%%%%%%%%%%%%%%%%%%%%%%%%%%%%%%%%%%%%%%%%%%%%%%%%%%%%%%%%%%%%%%
\subsection{The zero-bin subtraction and the universal TMDs}
\label{sec:zerobin}

Here we reorganize the factorization theorem such that the cross-section is expressed in terms of rapidity divergence-free TMDs as presented in the factorization theorem in \eqref{eq:FactGamGU}, \eqref{eq:FactGamGL} and \eqref{eq:FactGamFU}.
In this way we make clear the dependence on the universal TMDPDFs and  the unknown TMD soft function. To do this we write the TMD beam function in terms of the zero-bin un-subtracted term divided by the back-to-back soft function,
\begin{equation}\label{eq:ZBBeam}
\hat{B}_i(\xi,\bmat{b},\mu,k^- \delta_+) = \frac{ B_i^{\text{un.}} (\xi,\bmat{b},\mu, k^- / \delta^-) }{S (\bmat{b},\mu, \sqrt{\delta^+ \delta^{-}})}
\end{equation}
The back-to-back two-direction soft function operator definition can be found in appendix~\ref{sec:app-1}. 
Then following \cite{GarciaEchevarria:2011rb,Echevarria:2015byo} we can factorize the soft function as 
\begin{equation}
S (\bmat{b},\mu,\sqrt{\delta^+ \delta^{-}})  = S^{\frac{1}{2}}(\bmat{b},\mu,\delta^+  \nu)  S^{\frac{1}{2}}(\bmat{b},\mu,\delta^-/\nu )
\,,
\end{equation}
where $\nu$ is an arbitrary positive number which plays a bookkeeping role and will be removed from the final result, introducing this way a constraint on the product of rapidity scales. 
The bare function $\left(S^{\text{bare}}\right)^{\frac{1}{2}}$ is 
\begin{align}
\left(S^{\text{bare}}_i (\bmat{b}, \delta)\right)^{\frac{1}{2}}&= 1 + a_s C_i \lbc  
-\frac{2}{\epsilon^2} +\frac{4}{\epsilon} \ln \lp \frac{\sqrt{2} \,\delta   }{\mu}  \rp + 
 \ln (B\mu^2 e^{2\gamma_E} ) \lb  4 \ln \lp \frac{\sqrt{2}\, \delta   }{\mu}  \rp + \ln (B\mu^2 e^{2\gamma_E} ) \rb 
+\frac{\pi^2}{6}  \rbc \nl 
&+{\cal O}(a_s^2),
\end{align}
where here and in the rest of this manuscript we used the shorthand notation
\begin{align}
\label{eq:gamma_i}
\gamma_g &=  \frac{\beta_0}{2C_A} , \;& \gamma_q &= \frac{3}{2}, \; & C_f &= C_F  = \frac{N_c^2 -1}{2N_c} ,\; & C_g &= C_A =N_c
\,.
\end{align}
Thus we can now reorganize the beam and soft function matrix elements (denoted by the ``hat" notation) into a product of TMDs as they appear in the factorization theorem,
\begin{equation}
\label{eq:reorg}
\hat{B}_i( \xi,\bmat{b}, \mu, k^- \delta_+)   \hat{S}_{\gamma i} ( \bmat{b}, \mu, \sqrt{A_n}\,\delta_+)=   F_i (\xi, \bmat{b}, \mu, \zeta_1 )  \;S_{\gamma i} (\bmat{b},   \mu, \zeta_2) 
\end{equation}
where  the functions in the r.h.s. of \eqref{eq:reorg} are respectively
\begin{equation}
 F_i (\xi, \bmat{b}, \mu, \zeta_1 ) = \frac{ B_i^{\text{un.}} (\xi, \bmat{b}, \mu, k^- / \delta^-) } {S^{\frac{1}{2}} (\bmat{b},\mu,\delta^- / \nu)} \Bigg |_{\sqrt{2} \,k^- / \nu \to \sqrt {\zeta_1} } ,
\end{equation}
that is, the universal TMDPDF as defined in other observables such as Drell-Yan and semi-inclusive DIS and
\begin{equation}
\label{eq:soft-ratio}
S_{\gamma i} (\bmat{b}, \mu, \zeta_2)  = \frac{ \hat S_{\gamma i}  (\bmat{b},\mu,\sqrt{A_n} \,\delta^+) }   {S^{\frac{1}{2}} (\bmat{b},\mu,\delta^+ \nu)}  \Bigg |_{ \nu/ \sqrt{2 A_{n}}  \to \sqrt{\zeta_2}},
\end{equation}
that is the unknown soft function  now incorporated in a rapidity divergent-free ratio. We can thus eliminate the dependence on the arbitrary parameter $\nu$ by introducing the following constraint
\begin{equation}
\label{eq:zz}
\zeta_1 \, \zeta_2 = \frac{(k^-)^2}{A_n} =  \frac{\hat{u} \;\hat{t}}{\hat{s}} 
\end{equation}
where $\hat{s}$, $\hat{t}$, and $\hat{u}$ are the partonic Mandelstam variables and thus the combination $\zeta_1 \,\zeta_2$ is Lorentz invariant and in the Breit frame, or any other frame boosted along the proton direction,  we have: $\zeta_1 \,\zeta_2 = p_T^2$.
Notice that the procedure to obtain \eqref{eq:zz} is totally analogous to the one used in Drell-Yan or SIDIS~\cite{Collins:2011zzd,Echevarria:2012js}, TMD factorization theorem. In that case we have that $\zeta_{1,2}$ have both a square mass dimension and $\zeta_1\zeta_2=Q^4$, while in the present case  $\zeta_2$ is dimensionless quantity but $\zeta_1$, as usual, has dimensions of mass squared. The natural way to choose the values of $\zeta_1$ and $\zeta_2$ is 
\begin{align}
\zeta_1 =p_T^2,\;\quad \zeta_2  =1
\end{align}
This way we have the standard evolution for the TMDPDF up to the hard scale and the ratio of soft functions in \eqref{eq:soft-ratio} has no large rapidity logarithms and thus does not require evolution in rapidity.  

The renormalized soft function can then be written in terms of the renormalization kernel $Z^S$ and the bare soft function which is simply the ratio of the bare functions that appear in \eqref{eq:soft-ratio}, 
 \begin{equation}
 S_{\gamma i}^{\text{bare}}(\bmat{b}, \zeta_2) = Z_{\gamma i}^{S}(\bmat{b}, \mu, \zeta_2)  S_{\gamma i}(\bmat{b}, \mu, \zeta_2)
 \end{equation}
In the $\overline{\text{MS}}$ scheme  for the $(\gamma^* g)$ channel  we have 
\begin{multline}
S_{\gamma g} (\bmat{b},\mu,\zeta_2) =   1 + a_s\lbc 
 C_F \lb \frac{\pi^2}{3}  + 2 \ln^2 \lp \frac{B \mu^2 e^{2\gamma_E}}{ -A_{\bmat{b}}}\rp + 4 \text{Li}_2(1+A_{\bmat{b}})  \rb \\[5pt]
 +C_A \lb  - 2 \ln (B \mu^2 e^{2\gamma_E})   \ln \zeta_2  - \ln^2(-A_{\bmat{b}}) - \frac{\pi^2}{3} - 2  \text{Li}_2(1+A_{\bmat{b}}) \rb 
\rbc +{\cal O}(a_s^2),
\end{multline}
and for the ($\gamma^* f$) channel we have 
\begin{multline}
 S_{\gamma f} (\bmat{b},\mu,\zeta_2) = 1 + a_s \lbc
C_A \lb  \frac{\pi^2}{6}  +  \ln^2 \lp \frac{B \mu^2 e^{2\gamma_E}}{ -A_{\bmat{b}}}\rp + 2 \text{Li}_2(1+A_{\bmat{b}}) + 2\ln(B \mu^2 e^{2\gamma_E} )   \ln \frac{(n \cdot v_1) (\bmat{v}_2 \cdot \bmat{b})}{(n \cdot v_2) (\bmat{v}_1 \cdot \bmat{b})}   \rb \\[5pt]
+ C_F \ln(B \mu^2 e^{2\gamma_E} ) \lb \ln (B \mu^2 e^{2\gamma_E}) -2 \ln \zeta_2 + 2 \ln \lp\frac{2  (n \cdot v_2)}{(v_1\cdot v_2) (n \cdot v_2)} \rp  -\frac{\pi^2}{6} + 4\,\text{ln} (-i \,\bmat{v}_1 \cdot \hat{\bmat{b}}) \rb 
 \rbc +{\cal O}(a_s^2),
\end{multline}
Note that the imaginary terms cancel in the sum and after taking the Fourier transform in momentum space, resulting in a real cross-section, which we have checked explicitly up to the NLO contributions. The corresponding renormalization functions are 
\begin{equation}
Z^{S}_{\gamma g} (\bmat{b},\mu,\zeta_2) =   1 + a_s\lbc 
 C_F \lb \frac{4}{\epsilon^2} + \frac{4}{\epsilon} \ln \lp \frac{B \mu^2 \,e^{2\gamma_E}}{- A_{\bmat{b}}}\rp  \rb
 - C_A  \frac{2}{\epsilon} \ln \zeta_2  \rbc +{\cal O}(a_s^2),
\end{equation}
and 
\begin{multline}
Z^{S}_{\gamma f} (\bmat{b},\mu,\zeta_2) =   1 + a_s \lbc 
  C_A \lb \frac{2}{\epsilon^2} -\frac{2}{\epsilon} \lp \ln \frac{(n \cdot v_1) (\bmat{v}_2 \cdot \bmat{b})}{(n \cdot v_2) (\bmat{v}_1 \cdot \bmat{b})} \rp+\ln \lp \frac{B \mu^2 e^{2\gamma_E}}{-A_{\bmat{b}}}\rp    \rb \\[5pt]
  +\frac{2}{\epsilon}C_F \lb \ln (B \mu^2 e^{2\gamma_E}) -\ln \zeta_2 +\ln \lp\frac{2  (n \cdot v_2)}{(v_1\cdot v_2) (n \cdot v_2)} \rp  +  2 \,\text{ln}(-i  \,\bmat{v}_1 \cdot \hat{\bmat{b}} ) \rb
 \rbc +{\cal O}(a_s^2),
\end{multline} 
The soft anomalous dimension can be obtained from the renormalization functions as follows, 
\begin{align}
\gamma_{S_{\gamma i}} &= - \lp Z^{S}_{\gamma i} \rp^{-1} \frac{d}{d \ln \mu} Z^{S}_{\gamma_i} 
\end{align}
The one-loop results for the soft function anomalous dimensions are collected in the next section.
%%%%%%%%%%%%%%%%%%%%%%%%%%%%%%%%%%%%%%%%%%%%%%%%%%%%%%%%%%%%%%%%%%%%%%%%%%%%%%%%%%
\subsection{Consistency check}
Each element of the factorized cross-section has a factorization scale dependence and it satisfies a renormalization group equation, 
\begin{equation}
\label{eq:generic-rge}
\frac{d}{d\ln\mu} G(\mu) =  \gamma_G(\mu)\, G(\mu) 
\end{equation}
where $G$ runs over all the functions in the factorization theorem and $\gamma_G$ is the corresponding anomalous dimension.  On the other hand, the cross-section is renormalization group invariant. Therefore, as required by consistency, the sum of all anomalous dimensions of the terms appearing in the factorized cross-section must vanish. In  the impact parameter space, where the cross-section is written as a product of these functions, we have, 
\begin{equation}
\label{eq:gamma-cons}
(\gamma^*g)\text{-channel}\qquad
\gamma_{H_{\gamma g}} + \gamma_{S_{\gamma g}}+\gamma_{F_g}+2\gamma_{J_f}+\gamma_{\mathcal{C}_1}+\gamma_{\mathcal{C}_2}+\gamma_\alpha = 0,
\end{equation}
and 
\begin{equation}
\label{eq:gamma-cons-f}
(\gamma^*f)\text{-channel}\qquad
\gamma_{H_{\gamma f}} + \gamma_{S_{\gamma f}}+\gamma_{F_f}+\gamma_{J_f}+\gamma_{J_g}+\gamma_{\mathcal{C}_f}+\gamma_{\mathcal{C}_g}+\gamma_\alpha = 0.
\end{equation}
We write the perturbative expansion of these anomalous dimension as
\begin{equation}
\label{eq:AD-expand}
\gamma=\sum_{n=1} a_s^n \gamma^{[n]},
\end{equation}
with $a_s=\alpha_s/(4\pi)$.
For the two channels in the dijet process the relevant anomalous dimensions up to one-loop are,
\begin{align} 
\gamma_{H_{\gamma g} }^{[1]} &= 4 \lbc   C_F \lb \ln \lp  \frac{ \hat{s} ^2}{\mu^4} \rp  -2 \gamma_q   \rb  + C_A  \ln \lp \frac{\hat{t} \,\hat{u} } {\hat{s} \mu^2} \rp  \rbc \,, \nl
\gamma_{H_{\gamma f} }^{[1]}  & = 4 \lbc  C_F \lb \ln\lp  \frac{\hat{u}^2}{\mu^4}\rp  - 2 \gamma_q  \rb  + C_A   \ln \lp \frac{\hat{s} \, \hat{t}}{\hat{u}\, \mu^2} \rp   \rbc \,,\nl
\gamma_{S_{\gamma g}} ^{[1]}  &= 4 \lbc  - C_A \ln \zeta_2  +  2 C_F \lb \ln (B \mu^2 \,e^{2\gamma_E}) - \ln \hat{s} +\ln p_T^2 +\ln(4 c_{\bmat{b}}^2 )\rb \rbc\,, \nl
\gamma_{S_{\gamma f}} ^{[1]}   & =  4 \lbc  (C_F +C_A) \lb  \ln (B  \mu^2 e^{2\gamma_E})  -\ln \hat{s} +\ln  p_T^2  + \ln (4 c_{\bmat{b}}^2 ) \rb  + (C_F-C_A)\lb  \ln \lp \frac{\hat{t} }{ \hat{u}  }  \rp   -  \kappa(v_f) \rb- C_F \ln\zeta_2    \rbc   \nl
\gamma_{F_i}^{[1]}       &=4 C_i \lb - \ln\lp \frac{\zeta_1}{\mu^2} \rp + \gamma_i \rb \,,\nl
\gamma_{J_i}^{[1]}   &= 4 C_i \lb   -\ln \lp \frac{p_T^2}{\mu^2} \rp  -\ln R^2 + \gamma_i  \rb\,, \nl
\gamma_{\mathcal{C}_g}^{[1]}&= 4 C_A \lb   -  \ln \lp B \mu^2 \,e^{2\gamma_E} \rp  + \ln R^2 -\ln (4 c_{\bmat{b}}^2 )  + \kappa(v_g)\rb \,,\nl
\gamma_{\mathcal{C}_i}^{[1]}&= 4 C_F \lb   -  \ln \lp B \mu^2 \,e^{2\gamma_E} \rp  + \ln R^2 -\ln (4 c_{\bmat{b}}^2 )  + \kappa(v_i)\rb \,,\nonumber \\[8pt]
\gamma_\alpha^{[1]} &= - 4 C_A \gamma_g\,,
\end{align} 
The imaginary component in the soft and collinear-soft anomalous dimension is denoted by $\kappa(v_i)$ where 
\begin{align}
\kappa(v_f) = - \kappa(v_{\bar{f}}) = - \kappa(v_g) = i \pi\, \text{sign}( c_{\bmat{b}}).
\end{align}
These anomalous dimensions, except the soft function which we calculated here, can be found in \cite{Becher:2009th,Becher:2012xr,Chien:2020hzh,Hornig:2016ahz,Buffing:2018ggv,Echevarria:2015byo}. We also used \eqref{eq:notation-AB} to expand $A_{\bmat{b}}$ in the soft function anomalous dimension in terms of $\hat{s}$, $p_T$, and $c_{\bmat{b}}$. It is now easy to confirm the cancelation of the anomalous dimensions at $\mathcal{O}(\alpha_s)$ which also serves as confirmation of the factorization theorem at the same order. 

%%%%%%%%%%%%%%%%%%%%%%%%%%%%%%%%%%%%%%%%%%%%%%%%%%%%%%%%%%%%%%%%%%%%%%%%%%%%%%%%%%
\section{Heavy-meson pair imbalance}
\label{sec:3}
In this section we consider the process of heavy-meson pair production, $\ell + p \to \ell +H+\bar{H} +X$, in  the back-to-back limit for which the transverse momentum  imbalance $\bmat{r}_T$ is measured,
\begin{equation}
\bmat{r}_T = \bmat{p}_{T}^{H} + \bmat{p}_T^{\bar{H}}\;,
\end{equation}
where we use the notation $H$  for generic heavy meson and  $\bar{H}$ for the corresponding anti-particle. The imbalance is measured in the Breit frame and in the region sensitive to TMDs, i.e.,  $|\bmat{r}_T | \ll p_T^{H,\bar H}$, the two heavy mesons are fragmented near the kinematic end-point and carry most of the energy of the heavy quark coming from the hard process. 

In contrast to the dijet process, for the heavy-meson pair production we only need to consider the photon-gluon fusion channel. From this perspective the formalism is simpler, but on the other hand the mass of the heavy meson, $m_H$, introduces a new scale which we need to consider. In the case when the heavy mesons are highly boosted, i.e., $p_{T}^{H} \gg m_H$, the factorization is similar to the dijet production discussed above. The cross-section is then expressed in terms of the same hard, soft, and beam functions, but the production of the final state heavy mesons is described by a heavy-quark jet function, $J_{Q \to H}$~\cite{Jaffe:1993ie,Fickinger:2016rfd},
\begin{multline}
    \frac{d\sigma (\gamma^* g)}{dx d\eta_H d\eta_{\bar{H}} dp_T d\bmat{r}_T} = H^{\mu \nu} _{\gamma^*g\to Q \bar{Q} }(\hat{s},\hat{t}, \hat{u},\mu) \int \frac{d \bmat{b}}{(2\pi)^2}  \, \exp(i \bmat{b} \cdot \bmat{r}_T) \,F_{g, \mu \nu}(\xi, \bmat{b}, \mu,\zeta_1) \\
    \times S_{\gamma g}(\bmat{b},\mu,\zeta_2) \,J_{Q \to H}(\bmat{b}, p_T,m_Q, \mu) \,J_{\bar{Q} \to \bar{H}}(\bmat{b}, p_T, m_Q, \mu)\;.
\end{multline}
 with $\eta_H$ and $\eta_{\bar{H}}$ referring to the pseudo-rapidities of the heavy mesons.
Similarly to the dijet factorization, in the hard function we do not consider corrections due to the quark mass and we define, 
\begin{equation}
p_T  = \frac{|\bmat{p}_{T}^{H} | + |\bmat{p}_{T}^{\bar{H}}| }{2}\;,
\end{equation}
The decomposition into the unpolarized and linearly polarized gluon contributions follows the same steps as in section~\ref{sec:2} and thus we do not repeat here.

%%%%%%%%%%%%%%%%%%%%%%%%%%%%%%%%%%%%%%%%%%%%%%%%%%%%%%%%%%%%%%%%%%%%%%%%%%%%%%%%%%
\subsection{Refactorization of heavy-quark fragmentation function}

The fragmentation of a heavy quark to a heavy meson is described by the heavy-quark fragmentation function which is studied in a plethora of processes (see for example \cite{Zhu:2013yxa, Zhang:2017uiz, Boer:2010zf}). The large scale of the process, which is introduced  by the mass of the heavy quark, allows for the use of perturbation theory to calculate the fragmentation function up to small and universal non-perturbative corrections.  

In our case the heavy-quark jet function, $J_{Q\to H}$, describes the fragmentation of heavy mesons from heavy quarks and is differential in the two-dimensional transverse momentum of the fragments w.r.t. the beam axis. In the limit $r_T \ll p_T$ there are two parametrically different scales which are involved in the fragmentation process,  
\begin{align}\label{eq:hqscales}
\mu_+ = m_Q ,\;\quad  \text{ and } \;\quad  \mu_{\mathcal{J}} = m_Q \frac{r_T}{p_T}\;.
\end{align}
Logarithms of ratios of these scales will appear in the perturbative calculation of the jet function and could potentially ruin the convergence of perturbative expansion. Thus, resummation of these logarithms is  essential to ensure the convergence of the expansion. Note that these logarithms are the same to the  logarithms resummed by the TMD evolution.
The resummation  of the logs generated by the scales in \eqref{eq:hqscales} can be achieved through the means of factorization. To this end we employ the boosted-heavy-quark effective field theory (bHQET)~\cite{Fleming:2007qr}  which will allow us to factorize the jet function into a hard matching coefficient and a transverse momentum dependent matrix element. To demonstrate how such a factorization occurs, we give a brief description of the relevant modes. 

First consider the momentum of the heavy quark in the heavy meson rest frame, $p_Q^{\mu}$, which can be decomposed into  a mass term and the residual soft component, 
\begin{align}
\label{eq:rest}
p_Q^{\mu} \Bvert_{\text{rest frame}} &= m_Q \beta^{\mu} + k_s^{\mu}\,,& k_s^{\mu} &\sim \Lambda_{\text{QCD}}(1,1,1)_{v},
\end{align}
where $k_s^{\mu}$ is the typical size of soft (light) degrees of freedom in the heavy meson and $\beta^{\mu} = (1,1,0_{\perp})_v$. Note that here we decomposed four-vectors into the light-cone coordinates along the direction of the boosted heavy meson, $v$. The momenta $k_s^{\mu}$, in the boosted frame sets the size of energy loss  of the heavy quark during fragmentation to a heavy meson. To obtain the boosted momenta we simply apply the following transformations,
\begin{align}
\label{eq:boost}
v\cdot k &\to \Lambda v\cdot k, & \bar{v}\cdot k &\to \frac{\bar{v} \cdot k}{\Lambda}, & k_{\perp} & \to k_{\perp}.
\end{align} 
We can obtain $\Lambda$ by comparing the momentum of the heavy quark at the rest frame of the heavy meson in \eqref{eq:rest} to the momentum of the boosted heavy quark (up to power-corrections $\sim m_H/p_T^H)$, 
\begin{equation}
p_Q^{\mu} \Bvert_{\text{boosted frame}} \simeq \lp 2E_H , \frac{m_H^2}{2 E_H}, \Lambda_{\text{QCD}}\rp_v,
\end{equation}
and thus $\Lambda = 2E_H / m_H$. The transformations in \eqref{eq:boost} give the momentum scaling of the so called ``ultra-collinear" modes which are simply the  soft modes of  HQET boosted to the frame where we perform the measurement, 
\begin{equation}
k_{uc}^{\mu} \sim \Lambda_{\text{QCD}} \lp \frac{2E_H}{m_H},  \frac{m_H}{2E_H}, 1  \rp.
\end{equation}
The contribution to the  transverse momentum spectrum w.r.t. the beam axis, comes from the large component $\bar{v} \cdot k_{uc}$. Therefore, we can estimate  the typical size of the transverse momentum imbalance, $r_T \sim  \Lambda_{\text{QCD}} (2p_T^H /m_H)$. Since $p_T^H \gg m_H$, the typical soft scale, $r_T$, is  perturbative and that justifies the approach of perturbative matching to the TMD matrix elements to which non-perturbative effects are incorporated as corrections.

To proceed with the factorization we match from the massive-SCET~\cite{Fleming:2007qr,Fleming:2007xt}, which includes collinear degrees of freedom,  onto the boosted HQET where the degrees of freedom are the ultra-collinear modes. 
We are interested in the matching of the massive, collinear gauge invariant, quark building block, $\chi_{v} = W^{\dag}_v \xi_{v} $ 
with $$
W_v^\dagger(x)=\text{P}\exp\left(i g\int_0^\infty ds\; \bar n\cdot A_n (\bar n s+x)\right),$$
onto the HQET heavy quark fields, $h_{v\beta_+},$\footnote{In our notation $\beta^{\mu}$ is the collinear velocity of the heavy hadron and $v^{\mu}$ is the lightlike vector along the direction of the boosted quark. Thus we write $h_{v\beta_+}$ to indicate the typical velocity of the heavy quark expansion. The Wilson lines appearing on l.h.s. and r.h.s. of \eqref{eq:scettobhqet} are formally the same although the fields have a different scaling in the two cases.}
\begin{equation}\label{eq:scettobhqet}
 W_v^\dagger \,\xi_{v} \to C_+ (m_Q,\mu) W^{\dag}_{v} h_{v\beta_+},
\end{equation}    
where $C_+ (m_Q,\mu)$ is the short distance matching coefficient and $\beta_+$ denotes the heavy quark velocity in the boosted frame.
With this matching we can now factorize the jet function into a short distance matching coefficient and a bHQET matrix element that depends on the transverse momentum of the ultra-collinear fragments, 
\begin{align}\label{eq:heavyjetfact}
J_{Q \to H} (\bmat{b},p_T,m_Q,\mu) &= H_{+} (m_Q,\mu) \mathcal{J}_{Q \to H} \lp \bmat{b}, \frac{m_Q}{p_T}, \mu \rp,
\end{align}
where
 \begin{equation}
 H_{+} (m_Q,\mu) = | C_{+} (m_Q,\mu) |^2  \;.
\end{equation}
 The operator definition of the  two-dimensional shape function is
\begin{equation}
\label{eq:hqet-jet}
\mathcal{J}_{Q \to H} (\bmat{r}) = \frac{1}{ 2\, p^-_H \, N_C } \sum_X \langle 0| \delta^{(2)}\lp \bmat{r} -   i \bmat{v} \,(\bar{v} \cdot \partial \rp)  W_{v}^{\dag} h_{v\beta_+} | X H \rangle \langle X H |  \bar{h}_{v,\beta_+} W_{v}  \, \slashed{\bar v}| 0 \rangle ,
\end{equation}
where $\bmat{v}$ is a Euclidean,  two dimensional, transverse component of light-like four-vector $v^{\mu}$ pointing along the direction of the boosted heavy meson. The impact parameter space expression is obtained by simply taking the Fourier transform, 
\begin{equation}
\label{eq:FT-jet}
\mathcal{J}_{Q \to H}\lp \bmat{b}, \frac{m_Q}{p_T}, \mu \rp = \int d\bmat{r} \exp(i \bmat{b} \cdot \bmat{r}) \mathcal{J}_{Q \to H}(\bmat{r})\;.
\end{equation}

The hard matching coefficient is known up to two-loops but the jet function $\mathcal{J}_{Q \to H}(\bmat{r})$,  as defined above, appears here for the first time. However, as we discuss later in this section, this jet function is related at the operator level to the fragmentation shape function from \cite{Jaffe:1993ie,Fickinger:2016rfd} in the near-end-point limit, ($z_{H} \to 1$).  The one-loop hard function, $H_+$ is, 
\begin{equation}
\label{eq:h-nlo}
H_+ (m_Q,\mu) = 1+ \frac{\alpha_s }{4\pi}C_F \lbc \ln \lp\frac{\mu^2}{m_Q^2} \rp + \ln^2 \lp\frac{\mu^2}{m_Q^2} \rp + 8 + \frac{\pi^2}{6} \rbc ,
\end{equation}
and the corresponding anomalous dimension is 
\begin{equation}
\gamma_{+} = \frac{\alpha_s C_F}{\pi} \lbc  \frac{1}{2 }- \ln \lp \frac{m_Q^2}{\mu^2} \rp \rbc\;.
\end{equation}
In the following section we show the calculation of the bHQET matrix element, $\mathcal{J}_{Q \to H}$ at NLO and we use this result to derive the one-loop anomalous dimension. We demonstrate the consistency of anomalous dimensions for this process at NLO and we give an all order statement that connects the matrix element  $\mathcal{J}_{Q \to H}$ to the near-end-point fragmentation shape function for heavy mesons.

%%%%%%%%%%%%%%%%%%%%%%%%%%%%%%%%%%%%%%%%%%%%%%%%%%%%%%%%%%%%%%%%%%%%%%%%%%%%%%%%%%
\subsection{The  bHQET matrix element at NLO}
\begin{figure}
\begin{center}
\includegraphics[width=\textwidth]{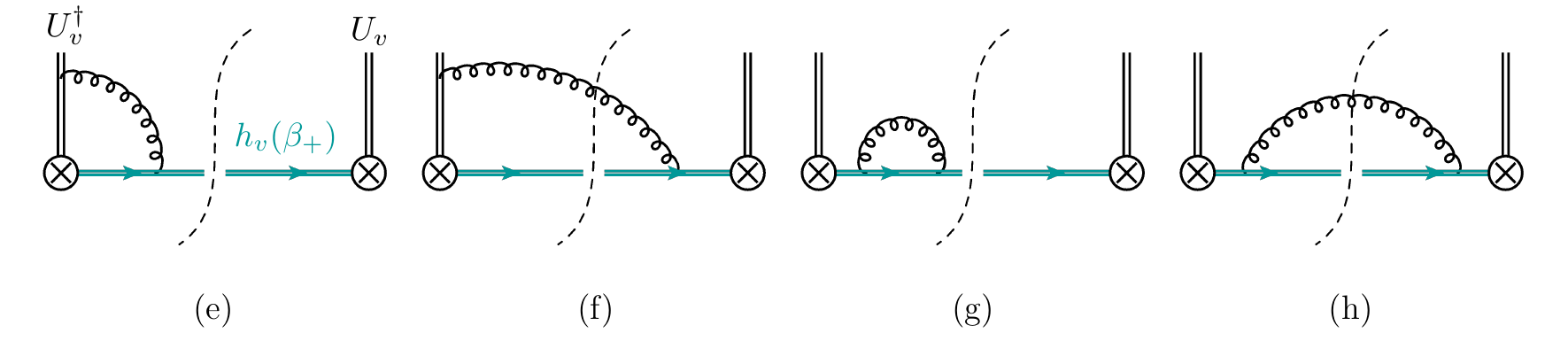}
\caption{\label{fig:bHQET} bHQET jet function at NLO diagrams. }
\end{center}
\end{figure}

The one-loop contributions to $\mathcal{J}_{Q \to H}(\bmat{b})$ in \eqref{eq:FT-jet} are shown in fig.~\ref{fig:bHQET}. We have non zero contributions only from diagrams (f) and (h) where a real gluon is exchanged. Virtual diagrams (e) and (g) are scaleless and vanish in our scheme,
\begin{equation}
\label{eq:hqet-jet-nlo-e}
\mathcal{J}_{Q \to Q} ^{\text{(f)+(h)}} =\frac{\alpha_s C_F}{\pi} (1-\epsilon)\,e^{ - \gamma_E \epsilon} \Gamma(\epsilon) \Gamma(-2\epsilon)  \,\mathcal{R}^{2 \epsilon} 
\end{equation}
where 
\begin{equation}
 \mathcal{R} = -  \frac{i\,p_T \mu \, e^{\gamma_E} (\bmat{v}\cdot \bmat{b} ) }{m_Q |\bmat{v}| } 
\end{equation}
where we have made the standard $\overline{MS}$ scale replacement $\mu^2 \to \mu^2 \exp(\gamma_E) /(4\pi)$. Expanding \eqref{eq:hqet-jet-nlo-e} in the limit $\epsilon \to 0$  and keeping all $\epsilon$ poles and finite non-vanishing terms we have 
\begin{equation}
\label{eq:jetB-nlo}
\mathcal{J}^{\text{bare}}_{Q \to Q} \lp \bmat{b}, \frac{m_Q}{p_T} \rp =  1+ \frac{\alpha_s C_F}{\pi} \lbc 
-\frac{1}{2\epsilon^2} + \frac{1}{2\epsilon} \lb 1- 2 \ln \mathcal{R} \rb +   \ln \mathcal{R} -  \ln^2 \mathcal{R} -\frac{5 \pi^2}{24} 
\rbc
\end{equation}
The renormalized jet function which in the $\overline{\text{MS}}$-scheme is simply given by the finite terms in \eqref{eq:jetB-nlo} is defined by the following equation,
\begin{equation}
\mathcal{J}^{\text{bare}}_{Q \to Q} \lp \bmat{b}, \frac{m_Q}{p_T} \rp = Z_{\mathcal{J}}  \lp \bmat{b}, \frac{m_Q}{p_T} \mu\rp\mathcal{J}_{Q \to Q} \lp \bmat{b}, \frac{m_Q}{p_T} , \mu\rp .
\end{equation} 
The corresponding anomalous dimension is 
\begin{equation}
\gamma_{\mathcal{J}} = - Z_{\mathcal{J}}^{-1} \frac{d}{d\ln\mu} Z_{\mathcal{J}} = \frac{\alpha_s C_F}{\pi} \lbc  1- 2 \ln \mathcal{R} \rbc
\end{equation}
It is now trivial to show the  consistency of the anomalous dimensions and therefore of factorization at NLO, since it is sufficient to show, 
\begin{equation}
\gamma_{\mathcal{J}} + \gamma_+ = \gamma_J + \gamma_{{\cal C}_f} .
\end{equation}
The result in \eqref{eq:jetB-nlo} involves logarithms of the scale $m_Q / (p_T b_0)$, where $b_0 = b \exp(\gamma_E) /2$. On the other hand the hard function $H_+$ in \eqref{eq:h-nlo} involves only logarithms of $m_Q$. This suggests that we have successfully separated the two scales and we can resum ratios of those (i.e., logarithms of $p_T b_0$) by evaluating each function at its canonical scale and then evolving each up to a common scale by solving with the corresponding renormalization group equation. 

%%%%%%%%%%%%%%%%%%%%%%%%%%%%%%%%%%%%%%%%%%%%%%%%%%%%%%%%%%%%%%%%%%%%%%%%%%%%%%%%%%
\subsection{Connection to the fragmentation shape function}

In this section we show that the two dimensional bHQET jet function in \eqref{eq:hqet-jet} is related to the fragmentation shape function. We consider the  operator definition of the shape function 
\begin{align}
S_{Q\to H}(\omega)=\frac{1}{2 N_c}\sum_X\langle 0 | \delta (\omega-i \sqrt{2} \,\bar v \cdot \partial )W_v^\dagger h_{v\beta_+} |H_{\beta} X\rangle\langle H_{\beta} X| \bar{h}_{v,\beta_+} W_v \frac{\slashed{\bar v}}{\sqrt{2}}|0\rangle,
\end{align}
as in eq.~(4.23) of \cite{Fickinger:2016rfd}. The same same shape function was  written with a  different notation in \cite{Jaffe:1993ie}.
 Taking the one dimensional Fourier transform of this expression with respect to $\omega$ we have\footnote{Note the normalization of the Hilbert states $| H \rangle = \sqrt{m_H} | H_{\beta} \rangle$} 
\begin{align}
\tilde{S}_{Q\to H} (\tau) &= \int d\omega \exp(i \omega \tau ) S_{Q \to H}(\omega) \nl &= 
\frac{1}{2\, m_H N_C } \sum_X \langle 0| \exp(- \sqrt{2}\,\tau \,\bar{v} \cdot \partial  ) \;W_{v}^{\dag} h_{v\beta_+} | X H \rangle \langle X H |  \bar{h}_{Qv} W_{v}  \frac{\slashed{\bar v}}{\sqrt{2}}|0 \rangle .
\end{align}
Comparing this result with the two dimensional Fourier transform of \eqref{eq:hqet-jet} as prescribed in \eqref{eq:FT-jet} we have 
\begin{equation}
\label{eq:jet-universality}
\mathcal{J}_{Q \to H}(\bmat{b}) = \frac{m_H}{\sqrt{2} \, p_{H}^-}\tilde{S}_{Q\to H} \lp \tau \to \frac{\bmat{v} \cdot \bmat{b}  } {\sqrt{2}} \rp .
\end{equation}
Using this equation we can confirm our perturbative calculation at NLO in \eqref{eq:jetB-nlo} by comparing the finite terms of this equation against  eq.~(4.24) of \cite{Fickinger:2016rfd}. 
To do that one needs the Fourier transformations of the regular plus-distributions which can be found in the literature but we also give here for completeness,
\begin{align}
\mathcal{FT}\lb \delta(\omega) \rb &= 1,\; &
\mathcal{FT}\lb \lp \frac{\Theta(\omega)}{\omega} \rp_+ \rb & = -\ln (-i \tau e^{\gamma_E}),\; &
\mathcal{FT}\lb \lp  \frac{ \Theta(\omega) \ln \omega }{\omega} \rp_+ \rb & = \frac{1}{2} \ln^2 (-i \tau e^{\gamma_E}) + \frac{\pi^2}{12}.
\end{align}
Using these equations we can easily check that indeed \eqref{eq:jet-universality} is satisfied up to NLO, although the perturbative validity of  \eqref{eq:jet-universality} is inferred beyond NLO. 
Therefore, since the anomalous dimension of the fragmentation shape function and the hard function $H_+$ is already known up to two-loops, we can use the consistency of factorization to solve for the anomalous dimension of the global soft function $S_{\gamma g}$ up to two-loops,
\begin{align}\label{eq:ADsoft2loops}
\gamma_{S_{\gamma g}}=-\left(\gamma_{H_{\gamma g}}+\gamma_{F_g}+\gamma_\alpha+\gamma_{{\cal J}}(\bmat{v}_1) + \gamma_{{\cal J}}(\bmat{v}_2)+2 \gamma_+ 
\right).
\end{align}
The anomalous dimensions at two and three loops are given in appendix \ref{sec:app-2}. 
We give the result for the soft function by organizing it into a term proportional to the cusp anomalous dimension, $\gamma_\text{cusp}$ and a ``non-cusp" term,
\begin{equation}\label{eq:SoftFuncAD}
\gamma_{S_{\gamma g}} = \gamma_{\text{cusp}}  \lb 2 C_F \ln \lp \frac{B\mu^2 e^{2 \gamma_E}}{-A_{\bmat{b} }  } \rp- C_A \ln \zeta_2 \rb +\delta \gamma_{S_{\gamma g}} ,
\end{equation}
where 
\begin{align}
\delta \gamma^{[1]}_{S_{\gamma g}}  &= 0\nl
\delta \gamma^{[2]}_{S_{\gamma g}} &= C_F \lb  C_A \lp \frac{1616}{27} -\frac{22}{9} \pi^2-56\, \zeta_3 \rp + n_fT_F \lp -\frac{448}{27} + \frac{8}{9} \pi^2 \rp \rb,
\end{align} 
and we have used the same notation for the perturbative expansion of the cusp anomalous dimension as in \eqref{eq:AD-expand}. 
The lengthy and not so intuitive three-loop non-cusp component, $\delta \gamma_{S_{\gamma g}}^{[3]}$, is given in the appendix (see \eqref{eq:3-loop-cusp}).

With this result we can push the calculation of the heavy-meson pair production up to NNLL with no additional perturbative calculations.  Furthermore with the knowledge of the  soft anomalous dimension and using \eqref{eq:gamma-cons} we can also solve for the collinear-soft anomalous dimension to the same order. 
This can now give us the NNLL cross-section of the dijet photon-gluon-fusion ($\gamma^* g$) process.
For the full NNLL dijet cross-section we are still missing the global-soft or collinear-soft anomalous dimensions from the photon-quark-initiated ($\gamma^* f$) process.

%%%%%%%%%%%%%%%%%%%%%%%%%%%%%%%%%%%%%%%%%%%%%%%%%%%%%%%%%%%%%%%%%%%%%%%%%%%%%%%%%%
\section{Conclusions}
\label{sec:4}

In this work we have established a new factorization theorem for dijet and heavy-meson pair production  in DIS which can be valuable in the quest of processes with a clear sensitivity to gluon TMDs. 
The factorization involves a new soft function, which we have calculated at one-loop, and whose anomalous dimension has been deduced at two and three loops from consistency relations. 
All the calculations have been performed with the $\delta$-regulator, combined with standard dimensional regularization. 
The factorized cross-section is then written terms of TMD parton distribution functions, the new TMD soft function and two final-state jet functions or heavy hadron distributions. 
The cross-section is sensitive to both unpolarized and linearly polarized gluon TMDs.

The influence of this new soft function is certainly an element that should be studied in the future. 
In particular one should understand how large is its non-perturbative contribution to the cross-section and whether it appears in multiple processes, i.e. whether it is a universal quantity. 
For a recent discussion on the universality of TMDs with multiple collinear directions see \cite{Boglione:2020cwn}.  

In our dijet analysis we do not consider the effects of any possible non-global logarithms which could be generated from in-out of jet correlations of the collinear-soft modes and are not associated with any of the TMD matrix elements (TMD-soft and TMD-PDF). 
Also we expect their effect to the resummed cross-section to be particularly small for the kinematic region of interest: $p_T \in[5,40]$~GeV and for small jet radius $R\sim 0.4$ ~\cite{Chien:2019gyf}. 
Thus, these effects could be incorporated into the ``jet-smearing" effects, from the hadronization of the jets, which are expected to be larger or of the same size. 
To this end, other possible extensions of this work can improve on this aspect by implementing modern jet substructure techniques, such as grooming, to reduce the sensitivity of the jets to non-global logarithms and hadronization effects. 
Recently in \cite{Chien:2020hzh} the angular de-correlation between a color-singlet boson ($\gamma, Z, W^{\mu}$) and  the winner-take-all (WTA) axis was studied in hadronic collisions, and it was shown to be free from non-global logarithms and, in addition, to have small sensitivity to the choice between charged-particles-only (tracks) or full jets. 
This last property of the angular de-correlation measurement using the WTA axis can be particularly useful when experimental limitations exist on the reconstruction of full jets. 
Therefore, extensions of \cite{Chien:2020hzh} in dijet process in DIS are of great interest to both jet and gluon TMD studies.  

This study focusses on the theoretical framework and the necessary elements for the resummed cross-section.
The details of a numerical study can depend on various aspects, such as the schemes for the TMD evolution and treatment of power corrections. 
We postpone a more quantitative analysis for a future study. 
In addition, a natural extension of this work is to incorporate spin effects from a polarized target hadron. 
Such effects will give rise to spin asymmetries, and particularly interesting is the case of the Sivers asymmetry (recently studied in hadronic dijet production~\cite{Kang:2020xez}). 
Thus, a simple generalization of our formalism can help formulate a factorization framework for processes with sensitivity to the gluon Sivers function.

\section*{Acknowledgements}
M.G.E., R.F.C. and I.S. are supported by the Spanish Ministry grant PID2019-106080GB-C21. This project has received funding from the European Union Horizon 2020 research and innovation program under grant agreement Num. 824093 (STRONG-2020). Y.M. is supported by the European Union’s Horizon 2020 research and innovation program under the Marie Sk\l{}odowska-Curie grant agreement No. 754496-FELLINI.

%%%%%%%%%%%%%%%%%%%%%%%%%%%%%%%%%%%%%%%%%%%%%%%%%%%%%%%%%%%%%%%%%%%%%%%%%%%%%%%%%%
\appendix
\section{Elements of factorization}
\label{sec:app-1}
In this section we list every function involved in the cross-section factorization for the dijet case as given in \eqref{eq:FactGamGU}, \eqref{eq:FactGamGL} and \eqref{eq:FactGamFU}. 
Additionally, we include the back-to-back two-direction soft function introduced in sec. \ref{sec:zerobin}.
%%%%%%%%%%%%%%%%%%%%%%%%%%%%%%%%%%%%%%%%%%%%%%%%%%%%%%%%%%%%%%%%%%%%%%%%%%%%%%%%%%
\subsection{Hard function}
The hard kernel can be found in \cite{Becher:2009th,Becher:2012xr,Chien:2020hzh}. The $\mu$-dependent part for both channels is given by
\begin{align}
    H^{U}_{\gamma^* g} &= 1 + a_s\left[-\left(2 C_{F}+C_{A}\right) \ln ^{2} \frac{\mu^{2}}{\hat{s}} + \right(2 C_{A} \ln \frac{\hat{t}\hat{u}}{\hat{s}^2} - 6 C_{F} \left) \ln \frac{\mu^{2}}{\hat{s}}+ \ldots \right] + \mathcal{O} (a_s^2),
\nl
    H^{U}_{\gamma^* f}& = 1 + a_s\left[-\left(2 C_{F} + C_A \right) \ln ^{2} \frac{\mu^{2}}{\hat{s}} + \left(2 C_{A} \ln \frac{\hat{t}}{\hat{u}}+4 C_{F} \ln \frac{-\hat{u}}{\hat{s}} - 6 C_{F}\right) \ln \frac{\mu^{2}}{\hat{s}} + \ldots \right] + \mathcal{O} (a_s^2),\nl
    H^{L}_{\gamma^* g} &=  1+ \mathcal{O} (a_s).
\end{align}
%%%%%%%%%%%%%%%%%%%%%%%%%%%%%%%%%%%%%%%%%%%%%%%%%%%%%%%%%%%%%%%%%%%%%%%%%%%%%%%%%%
\subsection{Jet function}
The definition of the jet function is as in \cite{Ellis:2010rwa},
\begin{align}
 J_{f, v}(v\cdot p,R)&=   \frac{1}{2 \sqrt{2} N_c}\text{Tr}\int d^4 x e^{i p x} \langle 0 |\bar \chi_v(p) \slashed{\bar v}\delta_{\text{alg}}(R) \chi_v(0)|0\rangle\\
\eta_\perp^{\rho\nu} J_{g, v}(v\cdot p,R)&=  - \frac{1}{ (N_c^2-1)}\sum_A\int d^4 x (\sqrt{2}\bar v \cdot p) e^{i p x}   \langle 0 |B^{\perp\rho A}_{v}(x) \delta_{\text{alg}}(R) gB^{\perp \nu A}_{v}(0)|0\rangle
 \end{align}
 where the symbol $\perp$ refers to the plane orthogonal to $v$  and whose  trasverse components are obtained with the tensor $\eta^\perp_{\alpha\beta} =g_{\alpha\beta}-(v_\alpha \bar v_\beta+\bar v_\alpha v_\beta)$. The perturbative calculation of the jet function can be found in \cite{Hornig:2016ahz}  and is given by
\begin{equation}
    J^{\mathrm{exc.}}_{i}(p_T,R,\mu)=1 + 2 a_s \lb C_i \lp\frac{1}{\epsilon^{2}} + \frac{\gamma_{i}}{\epsilon}\rp \lp \frac{\mu}{p_{T} R }\rp^{2 \epsilon}+d_{J}^{i, \mathrm{alg}}\rb + \mathcal{O} (a_s^2),
\end{equation}
The finite corrections $d^{i,\text{alg}}_J$ are given by
\begin{equation}
\begin{aligned}
d_{J}^{i, \text { cone }}= C_i \lp 2 \gamma_{i} \ln 2-\frac{5 \pi^{2}}{12} \rp +\left\{\begin{array}{ll}
C_{F} \frac{7}{2} & \text { if } i=q \\[5pt]
C_{A} \frac{137}{36}-T_{R} N_{f} \frac{23}{18} & \text { if } i=g
\end{array}\right.\\[5pt]
d_{J}^{i, k_{T}}=-C_{i} \frac{3 \pi^{2}}{4}+\left\{\begin{array}{ll}
C_{F} \frac{13}{2} & \text { if } i=q \\[5pt]
C_{A} \frac{67}{9}-T_{R} N_{f} \frac{23}{9} & \text { if } i=g
\end{array}\right.
\end{aligned}
\end{equation}
where $d_{J}^{i, k_{T}}$ is the same constant for all $k_{T}$-type algorithms $\left(k_{T}, \text { anti-} k_{T}, \text { and } \mathrm{C} / \mathrm{A}\right)$. 
\subsection{Collinear-soft function}
The collinear-soft function can be found in \cite{Buffing:2018ggv} and is given by the following matrix element, 
\begin{equation}
  \mathcal{C}_{i}(\bmat{b},R,\mu) = \int d\bmat{b} \exp( \bmat{b} \cdot \bmat{v}\, \bar{v} \cdot \partial) \frac{1}{N_R} \text{Tr} \langle 0|T \lb U_n^{\dagger}W_t (0)\rb \,\Theta_{\text{alg.}} \, \bar{T} \lb W_t^{\dagger} U_n (0) \rb |0\rangle \;,
\end{equation}
where $\bmat{v}$ is a Euclidean,  two dimensional, transverse component of light-like four-vector $v^{\mu}$ pointing along the direction of the jet. The normalization constant $N_R$ is simply the size of the representation for  SU($N_c$) of the $W_t$ and $U_n$ Wilson lines. For quark jets (fundamental  representation) we have $N_R =  N_c$ and for gluon jets (adjoint representation) we have $N_R = N_c^2 -1$.  The function $\Theta_{\text{alg.}}$ ensures that only contribution from outside the jet will contribute to the jet-imbalance. At NLO the bare collinear-soft function is given by
\begin{equation}
    \mathcal{C}_i^{\text{bare}} (\bmat{b},R)= 1 + 4 a_s C_i \frac{\exp (-\gamma_{E} \epsilon) \Gamma(-2 \epsilon)}{\epsilon \,\Gamma(1-\epsilon)}\lp - \frac{ i \mu e^{ \gamma_E} (\bmat{v} \cdot \bmat{b}) }{|\bmat{v}|\,R}\rp^{ 2 \epsilon} + \mathcal{O} (a_s^2).
\end{equation}
%

%%%%%%%%%%%%%%%%%%%%%%%%%%%%%%%%%%%%%%%%%%%%%%%%%%%%%%%%%%%%%%%%%%%%%%%%%%%%%%%%%%
\subsection{Beam function}
\label{app:BeamDef}
The quark, anti-quark and gluon beam functions are given in \cite{Echevarria:2016scs} by
\begin{align}
\hat{B}_{q} \left(x, \bmat{b}\right) &=\frac{1}{2} \sum_{X} \int \frac{d \xi^{+}}{2 \pi} e^{-i x p^{-} \xi^{+}}\left\{T\left[\bar{q}_{i} \tilde{W}_{n}^{T}\right]_{a}\left(\frac{\xi}{2}\right)|X\rangle \gamma_{i j}^{-}\langle X| \bar{T}\left[\tilde{W}_{n}^{T \dagger} q_{j}\right]_{a}\left(-\frac{\xi}{2}\right)\right\}, \nl
\hat{B}_{\bar{q}} \left(x, \bmat{b} \right) &=\frac{1}{2} \sum_{X} \int \frac{d \xi^{+}}{2 \pi} e^{-i x p^{-} \xi^{+}}\left\{T\left[\tilde{W}_{n}^{T \dagger} q_{j}\right]_{a}\left(\frac{\xi}{2}\right)|X\rangle \gamma_{i j}^{-}\langle X| \bar{T}\left[\bar{q}_{i} \tilde{W}_{n}^{T}\right]_{a}\left(-\frac{\xi}{2}\right)\right\}, \nl
\hat{B}_{g \mu\nu} \left(x, \bmat{b} \right) &=\frac{1}{x p^{-}} \sum_{X} \int \frac{d \xi^{+}}{2 \pi} e^{-i x p^{-} \xi^{+}}\left\{T\left[F_{-\mu} \tilde{W}_{n}^{T}\right]_{a}\left(\frac{\xi}{2}\right)|X\rangle\langle X| \bar{T}\left[\tilde{W}_{n}^{T \dagger} F_{-\nu}\right]_{a}\left(-\frac{\xi}{2}\right)\right\}\,,
\end{align}
where $\xi= ( \xi^{+}, 0^-, \bmat{b} )_n$. The repeated color indices $a$ ($a=1, \ldots, N_{c}$ for quarks and $a=1, \ldots, N_{c}^{2}-1$ for gluons) are summed up. The representations of the color $\mathrm{SU}(3)$ generators inside the Wilson lines are the same as the representation of the corresponding partons. The Wilson lines $\tilde{W}_{n}^{T}(x)$ are rooted at the coordinate $x$ and continue to the light-cone infinity along the vector $n,$ where it is connected by a transverse link to the transverse infinity (that is indicated by the superscript $T$). The TMDs are obtained from the beam functions $\hat B_i$ as explained in sec. \ref{sec:zerobin}.

%%%%%%%%%%%%%%%%%%%%%%%%%%%%%%%%%%%%%%%%%%%%%%%%%%%%%%%%%%%%%%%%%%%%%%%%%%%%%%%%%%
\subsection{Back-to-back two-direction soft function}
\label{app:2}

The back-to-back two-direction soft function introduced in \eqref{eq:ZBBeam} can be found in \cite{Echevarria:2015byo,Echevarria:2016scs}, where is defined as
\begin{equation}
S \left( \bmat{b},\mu, \delta^+ \delta^{-} \right)=\frac{\operatorname{Tr}}{N_{c}}\langle 0 |T\left[S_{n}^{T \dagger} \tilde{S}_{\bar{n}}^{T}\right]\left(0^{+}, 0^{-}, \boldsymbol{b} \right) \bar{T}\left[\tilde{S}_{\bar{n}}^{T \dagger} S_{n}^{T}\right](0) | 0 \rangle
\end{equation} 
where $S_{n}^{T}$ and $\tilde{S}_{\bar{n}}^{T}$ are soft Wilson lines as defined in \cite{Echevarria:2016scs}. Up to one-loop order, the two-direction soft function is given by (for $n\cdot \bar n=1$)
\begin{equation}
    S(\bmat{b},\mu,\delta^+ \delta^-) = 1  - 4 a_s C_i (B e^{\gamma_E}\mu^2)^{\epsilon} \Gamma(-\epsilon)
    \lb \text{ln}\lp \frac{B 2\delta^+ \delta^-}{e^{-2 \gamma_E}}\rp - \psi(-\epsilon) - \gamma_E \rb + \mathcal{O} (a_s^2),
\end{equation}
where $\psi$ is the polygamma function. 

%%%%%%%%%%%%%%%%%%%%%%%%%%%%%%%%%%%%%%%%%%%%%%%%%%%%%%%%%%%%%%%%%%%%%%%%%%%%%%%%%%%%%%%%
\section{Anomalous dimensions}
\label{sec:app-2}
In this appendix we collect from the literature the anomalous dimensions of the elements of the cross-section needed to obtain the dijet soft function anomalous dimension at two and three loops for the $\gamma^* g$-channel, as explained in \eqref{eq:ADsoft2loops}. 
We follow the standard procedure of separating  all anomalous dimensions into a  term proportional to the cusp anomalous dimension and a non-cusp term. We first give the three-loop cusp which applies to all functions and then proceed to give the non-cusp term for each function separately. The anomalous dimensions we give here are the ones that correspond to the renormalization group equation in \eqref{eq:generic-rge}.
%%%%%%%%%%%%%%%%%%%%%%%%%%%%%%%%%%%%%%%%%%%%%%%%%%%%%%%%%%%%%%%%%%%%%%%%%%%%%%%%%%%%%%%%
\subsection{The cusp anomalous dimension and $\beta$ function}
\begin{align}
\Gamma^i_{\mathrm{cusp}}\left(a_{s}\right)&=4C_i\sum_{n} \Gamma^{[n]}a_s^{n},\qquad
\gamma_{\text{cusp}}(\alpha_s)=\frac{1}{C_i}\Gamma^i_{\mathrm{cusp}}(a_s)=4 \sum_{n} \Gamma^{[n]}a_s^{n},
\end{align}
where
\begin{align}
\Gamma^{[1]} &=1, \nl \Gamma^{[2]}&=\left(\frac{67}{9}-\frac{\pi^{2}}{3}\right) C_{A}-\frac{20}{9} T_{F} n_{f}, \nl
\Gamma^{[3]} &=C_{A}^{2}\left(\frac{245}{6}-\frac{134 \pi^{2}}{27}+\frac{11 \pi^{4}}{45}+\frac{22}{3} \zeta_{3}\right)+C_{A} T_{F} n_{f}\left(-\frac{418}{27}+\frac{40 \pi^{2}}{27}-\frac{56}{3} \zeta_{3}\right) \nl
&+C_{F} T_{F} n_{f}\left(-\frac{55}{3}+16 \zeta_{3}\right)-\frac{16}{27} T_{F}^{2} n_{f}^{2}
\,,
\end{align}
and for $i = f,g$  we have $C_f = C_F$ and $C_g = C_A$. 
The $\beta$-function is given by
\begin{align}
\beta\left(a_{s}\right)&=-2 \alpha_{s}\sum_{n=1} \beta^{[n-1]}a_s^{n},
\end{align}
where
\begin{align}
\beta^{[0]}&\equiv \beta_0=\frac{11}{3} C_{A}-\frac{4}{3} T_{F} n_{f}, \nl
\beta^{[1]}&=\frac{34}{3} C_{A}^{2}-\frac{20}{3} C_{A} T_{F} n_{f}-4 C_{F} T_{F} n_{f}, \nl
\beta^{[2]}&=\frac{2857}{54} C_{A}^{2}+\left(2 C_{F}^{2}-\frac{205}{9} C_{F} C_{A}-\frac{1415}{27} C_{A}^{2}\right) T_{F} n_{f}+\left(\frac{44}{9} C_{F}+\frac{158}{27} C_{A}\right) T_{F}^{2} n_{f}^{2},\nl
\beta^{[3]}  &=
\frac{149753}{6} + 3564\zeta_3
- \left( \frac{1078361}{162} + \frac{6508}{27}\,\zeta_3 \right) n_f
+ \left( \frac{50065}{162} + \frac{6472}{81}\,\zeta_3 \right) n_f^2
+ \frac{1093}{729}\,n_f^3.
\end{align}
%%%%%%%%%%%%%%%%%%%%%%%%%%%%%%%%%%%%%%%%%%%%%%%%%%%%%%%%%%%%%%%%%%%%%%%%%%%%%%%%%%%%%%%%
\subsection{bHQET heavy-quark jet function}
As explained in \eqref{eq:heavyjetfact}, the heavy-quark jet function factorizes into a bHQET hard function $H_+$ and a bHQET jet function $\mathcal{J}_{Q \to Q}$. Their anomalous dimension are given by
\begin{align}
\gamma_{\mathcal{J}}&=-2 \Gamma_{\mathrm{cusp}}^q\ln \left(-i   \frac{p_T \mu e^{\gamma_{E} } (\bmat{v} \cdot \bmat{b}) }{m_Q | \bmat{v} |}\right)+2 \delta\gamma_{\mathcal{J}}.\nl
\gamma_+&=\Gamma_{\text {cusp}}^q \ln \frac{\mu^{2}}{m_{Q}^{2}}+2\delta \gamma_{+}
\end{align}
The non-cusp anomalous dimensions for the hard function and the jet function are known up to two-loops  \cite{Fickinger:2016rfd} and are given by
\begin{align}
\delta\gamma_{\mathcal{J}}^{[1]} &=2 C_{F}, \nl
\delta\gamma_{\mathcal{J}}^{[2]} &=-C_{F}\left[C_{A}\left(\frac{110}{27}+\frac{\pi^{2}}{18}-18 \zeta_3\right)+T_{F} n_{f}\left(\frac{8}{27}+\frac{2 \pi^{2}}{9}\right)\right], \nl
\delta\gamma_{+}^{[1]} &=C_{F}, \nl
\delta\gamma_{+}^{[2]} &=C_{F}\left[C_{F}\left(\frac{3}{2}-2 \pi^{2}+24 \zeta_3\right)+C_{A}\left(\frac{373}{54}+\frac{5}{2} \pi^{2}-30 \zeta_3\right)-T_{F} n_{f}\left(\frac{10}{27}+\frac{2}{3} \pi^{2}\right)\right].
\end{align}
Their sum is known up to three loops and is given by
\begin{align}
\delta\gamma_{+}^{[3]}+\delta\gamma_{\mathcal{J}}^{[3]}=& C_{F}^{3}\left(\frac{29}{2}+3 \pi^{2}+\frac{8 \pi^{4}}{5}+68 \zeta_3-\frac{16 \pi^{2} \zeta_3}{3}-240 \zeta_5\right) \nl
&+C_{A}^{2} C_{F}\left(-\frac{1657}{36}+\frac{2248 \pi^{2}}{81}-\frac{\pi^{4}}{18}-\frac{1552 \zeta_3}{9}+40 \zeta_5\right) \nl
&+C_{A} C_{F}^{2}\left(\frac{151}{4}-\frac{205 \pi^{2}}{9}-\frac{247 \pi^{4}}{135}+\frac{844 \zeta_3}{3}+\frac{8 \pi^{2} \zeta_3}{3}+120 \zeta_5\right) \nl
&+C_{A} C_{F} T_{F} n_{f}\left(40-\frac{1336 \pi^{2}}{81}+\frac{2 \pi^{4}}{45}+\frac{400 \zeta_3}{9}\right) \nl
&+C_{F}^{2} T_{F} n_{f}\left(-46+\frac{20 \pi^{2}}{9}+\frac{116 \pi^{4}}{135}-\frac{272 \zeta_3}{3}\right) \nl
&+C_{F} T_{F}^{2} n_{f}^{2}\left(-\frac{68}{9}+\frac{160 \pi^{2}}{81}-\frac{64 \zeta_3}{9}\right).
\end{align}
%%%%%%%%%%%%%%%%%%%%%%%%%%%%%%%%%%%%%%%%%%%%%%%%%%%%%%%%%%%%%%%%%%%%%%%%%%%%%%%%%%%%%%%%
\subsection{TMDPDF}
The universal TMDPDF anomalous dimension is given in \cite{Echevarria:2016scs} by
\begin{equation}
\gamma_F=\Gamma^i_{\mathrm{cusp}} \ln \frac{\mu^2}{\zeta_1} - \gamma_V^i .
\end{equation}
The non-cusp anomalous dimension for the TMDPDF is known up to three loops and is given by
\begin{align}
\gamma_V^{q[1]} &=-6 C_{F} \nl
\gamma_V^{q[2]} &=C_{F}^{2}\left(-3+4 \pi^{2}-48 \zeta_{3}\right)+C_{F} C_{A}\left(-\frac{961}{27}-\frac{11 \pi^{2}}{3}+52 \zeta_{3}\right)+C_{F} T_{F} n_{f}\left(\frac{260}{27}+\frac{4 \pi^{2}}{3}\right) \nl
\gamma_V^{q[3]} &=C_{F}^{3}\left(-29-6 \pi^{2}-\frac{16 \pi^{4}}{5}-136 \zeta_{3}+\frac{32 \pi^{2}}{3} \zeta_{3}+480 \zeta_{5}\right) \nl
&+C_{F}^{2} C_{A}\left(-\frac{151}{2}+\frac{410 \pi^{2}}{9}+\frac{494 \pi^{4}}{135}-\frac{1688}{3} \zeta_{3}-\frac{16 \pi^{2}}{3} \zeta_{3}-240 \zeta_{5}\right) \nl
&+C_{F} C_{A}^{2}\left(-\frac{139345}{1458}-\frac{7163 \pi^{2}}{243}-\frac{83 \pi^{4}}{45}+\frac{7052}{9} \zeta_{3}-\frac{88 \pi^{2}}{9} \zeta_{3}-272 \zeta_{5}\right) \nl
&+C_{F}^{2} T_{F} n_{f}\left(\frac{5906}{27}-\frac{52 \pi^{2}}{9}-\frac{56 \pi^{4}}{27}+\frac{1024}{9} \zeta_{3}\right) \nl
&+C_{F} C_{A} T_{F} n_{f}\left(-\frac{34636}{729}+\frac{5188 \pi^{2}}{243}+\frac{44 \pi^{4}}{45}-\frac{3856}{27} \zeta_{3}\right)+C_{F} T_{F}^{2} n_{f}^{2}\left(\frac{19336}{729}-\frac{80 \pi^{2}}{27}-\frac{64}{27} \zeta_{3}\right)
\end{align}
\begin{align}
\gamma_V^{g[1]} &=-\frac{22}{3} C_{A}+\frac{8}{3} T_{F} n_{f}, \nl
\gamma_V^{g[2]} &=C_{A}^{2}\left(-\frac{1384}{27}+\frac{11 \pi^{2}}{9}+4 \zeta_{3}\right)+C_{A} T_{F} n_{f}\left(\frac{512}{27}-\frac{4 \pi^{2}}{9}\right)+8 C_{F} T_{F} n_{f}, \nl
\gamma_V^{g[3]} &=2 C_{A}^{3}\left(\frac{-97186}{729}+\frac{6109}{486} \pi^{2}-\frac{319}{270} \pi^{4}+\frac{122}{3} \zeta_{3}-\frac{20}{9} \pi^{2} \zeta_{3}-16 \zeta_{5}\right) \nl
&+2 C_{A}^{2} T_{F} n_{f}\left(\frac{30715}{729}-\frac{1198}{243} \pi^{2}+\frac{82}{135} \pi^{4}+\frac{712}{27} \zeta_{3}\right) \nl
&+2 C_{A} C_{F} T_{F} n_{f}\left(\frac{2434}{27}-\frac{2}{3} \pi^{2}-\frac{8}{45} \pi^{4}-\frac{304}{9} \zeta_{3}\right)-4 C_{F}^{2} T_{F} n_{f} \nl
&+2 C_{A} T_{F}^{2} n_{f}^{2}\left(-\frac{538}{729}+\frac{40}{81} \pi^{2}-\frac{224}{27} \zeta_{3}\right)-\frac{88}{9} C_{F} T_{F}^{2} n_{f}^{2}.
\end{align}
%%%%%%%%%%%%%%%%%%%%%%%%%%%%%%%%%%%%%%%%%%%%%%%%%%%%%%%%%%%%%%%%%%%%%%%%%%%%%%%%%%%%%%%%
\subsection{Hard Function}
The hard function $H_{\gamma^* g}$ anomalous dimension is given in \cite{Becher:2009th} by
\begin{equation}
\gamma_H=(2 C_{F}+C_{A}) \gamma_{\mathrm{cusp}} \ln \frac{p_{T}^{2}}{\mu^{2}}-2C_F\gamma_{\mathrm{cusp}} \ln \frac{\hat{t}\hat{u}}{\hat{s}^2} +
 \delta\gamma_{H}-\frac{\beta ( \alpha_{s}) }{\alpha_{s}}
\end{equation}
The non-cusp anomalous dimensions is given by
\begin{align}
\delta\gamma_{H}^{[1]}=&-2\beta_{0}-12 C_{F}, \nl
\delta\gamma_{H}^{[2]}=&\left(\frac{256}{27}-\frac{2 \pi^{2}}{9}\right) 2 C_{A} T_F n_{f} +\left(\frac{368}{27}+\frac{4 \pi^{2}}{3}\right) 2 C_{F} T_F n_{f} +\left(-3+4 \pi^{2}-48 \zeta_{3}\right) 2 C_{F}^{2} \nl
+&\left(-\frac{692}{27}+\frac{11 \pi^{2}}{18}+2 \zeta_{3}\right) 2 C_{A}^{2}+\left(-\frac{961}{27}-\frac{11 \pi^{2}}{3}+52 \zeta_{3}\right) 2 C_{A} C_{F}, \nl
\delta\gamma_{H}^{[3]}=&\gamma^{q[3]}_V + \frac{1}{2}\gamma^{g[3]}_V.
\end{align}

%%%%%%%%%%%%%%%%%%%%%%%%%%%%%%%%%%%%%%%%%%%%%%%%%%%%%%%%%%%%%%%%%%%%%%%%%%%%%%%%%%
\subsection{Dijet soft function}
The dijet anomalous dimension for the $\gamma^* g$-channel is given by eq.~(\ref{eq:SoftFuncAD}). The non-cusp three-loop anomalous dimension is given by
\begin{align}
\label{eq:3-loop-cusp}
\delta \gamma^{[3]}_{S_{\gamma g}} & = \frac{1}{7290} \lbc 8 C_F n_f T_F \lb 27 \left(2280 \zeta _3+12 \pi ^4+45 \pi ^2-4480\right) C_F-5 \left(-3024 \zeta _3+180 \pi ^2+4753\right) n_f T_F \rb \nl
& + 4 C_A n_f T_F \lb 5 \left(-3024 \zeta _3+180 \pi ^2-4535\right) n_f T_F + \left(36720 \zeta _3-2268 \pi ^4+19995 \pi ^2+188110\right) C_F \rb \nl
& - C_A^3 \lb 15 \pi ^2 \left(1080 \zeta _3-6109\right)+20 \left(-14823 \zeta _3+5832 \zeta _5+48593\right)+8613 \pi ^4 \rb \nl
& + 2 C_A^2 \lb -385695 + \left(96120 \zeta _3+2214 \pi ^4-17970 \pi ^2+535625\right) n_f T_F  \nl
& - \left(-1598940 \zeta _3+33 \pi ^2 \left(1080 \zeta _3+216 \pi ^2-2875\right)+699840 \zeta _5+683905\right) C_F \rb \rbc.
\end{align}

%%%%%%%%%%%%%%%%%%%%%%%%%%%%%%%%%%%%%%%%%%%%%%%%%%%%%%%%%%%%%%%%%%%%%%%%%%%%%%%%%%
\section{Dijet soft function integrals}
\label{sec:app-3}
In this section we give a pedagogical review of the integrals needed to obtain the dijet soft function result at one-loop order. The integrals are introduced in \eqref{eq:softintegrals}. 
The real diagrams are shown in fig.~\ref{fig:soft} and are the only ones contributing to the soft function. The virtual diagrams vanish, as we show in this section, and are not shown in this work. In the following, we use $d = 4 - 2\epsilon$.

%%%%%%%%%%%%%%%%%%%%%%%%%%%%%%%%%%%%%%%%%%%%%%%%%%%%%%%%%%%%%%%%%%%%%%%%%%%%%%%%%%
\subsubsection*{Real diagram $\bmat{n}$ - $\bmat{v_J}$}

The integral we need to compute is given by the expression
\begin{equation}
    I_{JB} = \int\frac{d^d k}{(2\pi)^d}\frac{e^{i \bmat{k}\cdot \bmat{b}}\delta(k^2)\theta(k_+)}{(n\cdot k+i\delta^+)(v_J\cdot k)}.
\end{equation}
Using the delta to integrate over $k_-$ we get
\begin{equation}
    I_{JB} =  \frac{1}{2\pi}\int\frac{d k_+}{2\pi}\int\frac{d^{d-2} \bmat{k}}{(2\pi)^{d-2}}\frac{e^{i \bmat{k}\cdot \bmat{b}}\theta(k_+)}{(k_+ + i\delta^+)(2v^{-}_J k^2_+ + v^{+}_J \bmat{k}^2 - 2k_+ \bmat{v}_J\cdot \bmat{k})}.
\end{equation}
Completing the square, the denominator can be written as
\begin{equation}
\begin{gathered}
    2v_J^- k^2_+ + v_J^+ \bmat{k}^2 - 2k_+\bmat{v}_J\cdot \bmat{k} = v_J^+ (\bmat{k}-\frac{k_+}{v_J^+}\bmat{v}_J)^2.
\end{gathered}
\end{equation}
We change variables $\bmat{k}' \rightarrow \bmat{k}-\frac{k_+}{v_J^+}\bmat{v}_J$. 
In this way, the integral simplifies to the following expression,
\begin{equation}\label{v2A}
   I_{JB} =   \frac{1}{2\pi v_J^+}\int\frac{d k_+}{2\pi}\frac{\theta(k_+)e^{i\frac{\bmat{v}_J\cdot \bmat{b}}{v^+_J}k_+}}{k_+ + i\delta^+}\int\frac{d^{d-2} \bmat{k}}{(2\pi)^{d-2}}\frac{e^{i \bmat{k}\cdot \bmat{b}}}{\bmat{k}^2}.
\end{equation}
This allows us to perform each integral separately, which leads us to the final result of the integral,
\begin{equation}
    I_{JB} = 
    -\frac{4B^\epsilon\Gamma(-\epsilon)}{(4\pi)^{3-\epsilon}v_J^+}\lb \ln \lp \frac{(\bmat{v}_J\cdot \bmat{b})\delta^+}{v_J^+}\rp +\gamma_E\rb.
\end{equation}

%%%%%%%%%%%%%%%%%%%%%%%%%%%%%%%%%%%%%%%%%%%%%%%%%%%%%%%%%%%%%%%%%%%%%%%%%%%%%%%%%%
\subsubsection*{Real diagram $\bmat{v_1}$ - $\bmat{v_2}$}
The integral we need to compute is given by
\begin{equation}
    I_{12} = \int\frac{d^d k}{(2\pi)^d}\frac{e^{i \bmat{k} \cdot \bmat{b}} \delta(k^2)\theta(k_+)}{(v_1 \cdot k)(v_2 \cdot k)}.
\end{equation}
Using the delta to integrate over $k_-$ we get 
\begin{equation}
\begin{gathered}
    I_{12} = \frac{1}{2\pi}\int\frac{d k_+}{2\pi}\int\frac{d^{d-2} \bmat{k}}{(2\pi)^{d-2}}\frac{e^{i \bmat{k} \cdot \bmat{b}}\theta(k_+)2k_+}{(2v_1^- k^2_+ + v_1^+ \bmat{k}^2 - 2k_+ \bmat{v}_1 \cdot \bmat{k})(2 v_2^- k^2_+ + v_2^+ \bmat{k}^2 - 2k_+ \bmat{v}_2 \cdot \bmat{k})}.
\end{gathered}
\end{equation}
We use Feynman parametrization in order to rewrite the denominator,
\begin{equation}
    \frac{1}{AB} = \int^1_0 dx \frac{1}{[Ax+(1-x)B]^2}.
\end{equation}
In our case, we identify
\begin{align}
         A &= \frac{2v_1^- k^2_+}{v_1^+} + \bmat{k}^2 - \frac{2k_+ \bmat{v}_1 \cdot \bmat{k}}{v_1^+},  \; &
         B &= \frac{2v_2^- k^2_+}{v_2^+} + \bmat{k}^2 - \frac{2k_+ \bmat{v}_2 \cdot \bmat{k}}{v_2^+}.
\end{align}
In this way, the denominator can be rewritten the following way:
\begin{equation}
         xA+(1-x)B = [\bmat{k} - k_+ \bmat{R}_1(x)]^2+k^2_+R_2(x),
\end{equation}
where
\begin{align}
    \bmat{R}_1(x) &= \frac{x\bmat{v}_1}{v_1^+}+\frac{(1-x)\bmat{v}_2}{v_2^+}, \;&
    R_2(x) &= 2x(1-x)\frac{v_1 \cdot v_2}{v_1^ + v_2^+}.
\end{align}
We can perform a change of variables $\bmat{k}' = \bmat{k} - k_+ \bmat{R}_1^\mu(x)$ and the integral simplifies as follows,
\begin{equation}
    I_{12} = \frac{1}{2\pi v_1^+v_2^+}\int^1_0dx\int\frac{d k_+}{2\pi}\int\frac{d^{d-2} \bmat{k}}{(2\pi)^{d-2}}\frac{e^{i \bmat{k} \cdot \bmat{b}}e^{i k_+ \bmat{R}_1 \cdot \bmat{b}}\theta(k_+)2k_+}{[\bmat{k}^2 + R_2(x)k^2_+]^2}.
\end{equation}
We perform a Mellin-Barnes transformation in order to be able to integrate over $\bmat{k}$ and $k_+$ separately,
\begin{equation}
    \frac{1}{[\bmat{k}^2+R_2k^2_+]^2}= \int^{+i\infty}_{-i\infty}\frac{dz}{2\pi i}\Gamma(2+z)\Gamma(-z)\frac{(R_2k^2_+)^z}{(\bmat{k}^2)^{2+z}}.
\end{equation}
In this way, we integrate over $\bmat{k}$ and get
\begin{equation}
    I_{12} = \frac{4}{(4\pi)^{2-\epsilon}v_1^+v_2^+} B^{\epsilon+1}\int^1_0dx\int^{+i\infty}_{-i\infty}\frac{dz}{2\pi i}\Gamma(-z)\Gamma(-\epsilon-1-z)B^z R^z_2\int^{+\infty}_0\frac{dk_+}{2\pi}k^{2z+1}_+e^{ik_+ \bmat{R}_1 \cdot \bmat{b}}.
\end{equation}
Next, we integrate over $k_+$,
\begin{equation}
    I_{12} = \frac{8}{(4\pi)^{3-\epsilon}v_1^+v_2^+} B^{\epsilon+1}\int^1_0dx\int^{+i\infty}_{-i\infty}\frac{dz}{2\pi i}\Gamma(-z)\Gamma(-\epsilon-1-z)\Gamma(2+2z)B^z(-i\bmat{R}_1 \cdot \bmat{b})^{-2(1+z)}R^z_2 .
\end{equation}
We integrate in $z$ using residues. We close the integration path to the right side of the imaginary axis and sum all the residues due to $\Gamma(-z)$ and $\Gamma(-\epsilon-1-z)$. The residues are given by the expressions
\begin{equation}
    \operatorname{Res}[\Gamma(-z), n]=\frac{(-1)^{n+1}}{\Gamma(n+1)}\qquad \text{and}\qquad \operatorname{Res}[\Gamma(-1-\epsilon-z), n-1-\epsilon]=\frac{(-1)^{n+1}}{\Gamma(n+1)},
\end{equation}
for $n = 0, 1, 2, ...$ We then arrive at the sum
\begin{equation*}
\begin{gathered}
    I_{12} =  \frac{-8}{(4\pi)^{3-\epsilon}v_1^+v_2^+}B^{\epsilon + 1}\int_{0}^{1} d x \sum^\infty_{n=0} \lbc \frac{\Gamma(-1-\epsilon-n) \Gamma(2+2 n)}{\Gamma(n+1)} \frac{\left(B R_{2}\right)^{n}}{(\bmat{R}^2_{1})^{n+1}} \\
     -\frac{\Gamma(1+\epsilon-n) \Gamma(-2 \epsilon+2 n)}{\Gamma(n+1)}(-1)^{\epsilon} \frac{\left(B R_{2}\right)^{n-1-\epsilon}}{(\bmat{R}^2_{1})^{n- \epsilon}} \rbc,
\end{gathered}
\end{equation*}
Now, we perform the $x$ integral, where remember that the $x$ dependence is through the $\bmat{R}_1(x)$ and $R_2(x)$ functions.
Finally, we perform the sum over $n$ and arrive at the final result:
\begin{equation}
\begin{gathered}
    I_{12} = \frac{4}{(4 \pi)^{3-\epsilon}(v_1 \cdot v_2)}\, B^{\epsilon} \,A_{\bmat{b}}\lb -\Gamma(-1-\epsilon)_{2} F_{1}(1,1,2+\epsilon,-A_{\bmat{b}}) + \Gamma^{2}(-\epsilon) \Gamma(1+\epsilon)(-1)^{\epsilon} A_{\bmat{b}}^{-1-\epsilon}(1+A_{\bmat{b}})^{\epsilon} \rb.
\end{gathered}
\end{equation}
where $A_{\bmat{b}}$ is given in \eqref{eq:notation-AB}.

%%%%%%%%%%%%%%%%%%%%%%%%%%%%%%%%%%%%%%%%%%%%%%%%%%%%%%%%%%%%%%%%%%%%%%%%%%%%%%%%%%
\subsubsection*{Virtual diagram $\bmat{n}$ - $\bmat{v_J}$}

The integral for this diagram is given by
\begin{equation}
    I^{\mathrm{virtual}}_{JB} = \int\frac{d^d k}{(2\pi)^d}\frac{1}{(n\cdot k+i\delta^+)(\bmat{v}_J \cdot \bmat{k}) \bmat{k}^2}.
\end{equation}
We begin integrating over $k_-$. 
This integration restricts $k_+$ values to be negative ($k_+<0$), otherwise all poles would be located in the negative part of the imaginary axis (due to $v_J^+ > 0$), and the integral would vanish. 
In this way, we get
\begin{equation}
    I^{\mathrm{virtual}}_{JB} =  
    i\int\frac{d k_+}{2\pi}\int\frac{d^{d-2} \bmat{k}}{(2\pi)^{d-2}}\frac{\theta(-k_+)}{(k_+ + i\delta^+)(2v_J^- k^2_+ + v_J^+ \bmat{k}^2 - 2k_+\bmat{v}_J \cdot \bmat{k})}.
\end{equation}
Completing the square and performing the same change of variable as in the real diagram $n-v_J$, we get
\begin{equation}
I^{\mathrm{virtual}}_{JB} = 
\frac{i}{v_J^+}\int\frac{d k_+}{2\pi}\frac{\theta(-k_+)}{k_+ + i\delta^+}\int\frac{d^{d-2} \bmat{k}}{(2\pi)^{d-2}}\frac{1}{\bmat{k}^2} = 0.
\end{equation}
This integral vanishes due to the integral over $\bmat{k}$ being a scaleless integral, which is set to zero in dimensional regularization. This was expected as the result of real diagram $n-v_J$ has no IR divergent part. 

%%%%%%%%%%%%%%%%%%%%%%%%%%%%%%%%%%%%%%%%%%%%%%%%%%%%%%%%%%%%%%%%%%%%%%%%%%%%%%%%%%
\subsubsection*{Virtual diagram $\bmat{v_1}$ - $\bmat{v_2}$}
The integral that has to be computed is given by
\begin{equation}
    I^{\mathrm{virtual}}_{12} = \int\frac{d^d k}{(2\pi)^d}\frac{1}{(v_1\cdot k+i0)(v_2\cdot k-i0)(k^2+i0)} .
\end{equation}
First, we integrate over $k_-$. We get a different result depending on the sign of $k_+$, which leads to the addition of $\theta(k_+)$ and $\theta(-k_+)$,
\begin{equation}
    \mathcal{I}_{1} = \int\frac{d k_+}{2\pi}\int\frac{d^{d-2} \bmat{k}}{(2\pi)^{d-2}}\frac{i\theta(-k_+)}{\lb \frac{2 v_1^-}{v_1^+} k^2_+ + \bmat{k}^2 - \frac{2k_+\bmat{v}_1 \cdot \bmat{k}}{v_1^+} \rb \lb \bmat{k} \cdot (\bmat{v}_2 v_1^+ - \bmat{v}_1 v_2^+) + k_+(v_2^+ v_1^- - v_2^- v_1^+) \rb},
\end{equation}
\begin{equation}
    \mathcal{I}_{2} = \int\frac{d k_+}{2\pi}\int\frac{d^{d-2} \bmat{k}}{(2\pi)^{d-2}}\frac{i\theta(k_+)}{\lb \frac{2 v_2^-}{v_2^+} k^2_+ + \bmat{k}^2 - \frac{2k_+\bmat{v}_2 \cdot \bmat{k}}{v_2^+} \rb \lb \bmat{k} \cdot (\bmat{v}_2 v_1^+ - \bmat{v}_1 v_2^+) + k_+(v_2^+ v_1^- - v_2^- v_1^+) \rb},
\end{equation}
\begin{equation}
    I^{\mathrm{virtual}}_{12} = \mathcal{I}_{1} + \mathcal{I}_{2}.
\end{equation}
Both integrals are computed the same way, so we focus in one of them, $\mathcal{I}_{1}$ for example. Here, we use Feynman parametrization in order to rewrite the denominator,
\begin{equation}
    \frac{1}{AB} = \int^1_0 dx \frac{1}{[Ax+(1-x)B]^2}.
\end{equation}
In our case, we identify
\begin{align}
        A &= \frac{2 v_1^-}{v_1^+} k^2_+ + \bmat{k}^2 - \frac{2k_+\bmat{v}_1 \cdot \bmat{k}}{v_1^+}, \;&
        B &= \bmat{k} \cdot (\bmat{v}_2 v_1^+ - \bmat{v}_1 v_2^+) + k_+(v_2^+ v_1^- - v_2^- v_1^+).
\end{align}
In this way, the denominator can be rewritten the following way:
\begin{equation}
        xA+(1-x)B =  R_1(x)[\bmat{k}-k_+\bmat{E}_1(x)-\bmat{E}_2 (x)]^2+k_+F_1(x) + F_2(x),
\end{equation}
where
\begin{equation}
    \begin{gathered}
        R_1(x) = x,\\
        \bmat{E}_1(x) = \frac{\bmat{v_1}}{v_1^+},\quad \bmat{E}_2(x) = \frac{1}{2} \left(\frac{v_2^+ \bmat{v}_1(1-x)}{x} - \frac{v_1^+ \bmat{v}_2 (1-x)}{x}\right),\\
        F_1(x) = -(v_1 \cdot v_2)(1-x)\quad \text{and}\quad F_2(x) = -(v_1 \cdot v_2)\frac{v_2^+ v_1^+ (1-x)^2}{2 x}.
    \end{gathered}
\end{equation}
We can perform a change of variables $\bmat{k}' = \bmat{k}-k_+\bmat{E}_1(x)-\bmat{E}_2 (x)$ and the integral simplifies to
\begin{equation}
    \mathcal{I}_{1} = \int^1_0dx\int\frac{d k_+}{2\pi}\int\frac{d^{d-2} \bmat{k}}{(2\pi)^{d-2}}\frac{i\theta(-k_+)}{[R_1(x)\bmat{k}^2 + k_+F_1(x) + F_2(x)]^2}.
\end{equation}
We can integrate over $\bmat{k}$ we get
\begin{equation}
    \mathcal{I}_{1} = \frac{i}{(4\pi)^{1-\epsilon}}\frac{\Gamma(\epsilon+1)}{\Gamma(2)}\int^1_0dx \int\frac{d k_+}{2\pi}\theta(-k_+)[k_+F_1(x) + F_2(x)]^{-\epsilon-1}\frac{1}{(R_1(x))^{1-\epsilon}}.
\end{equation}
Finally, integrating over $k_+$ we have
\begin{equation}
    \mathcal{I}_{1} = \frac{-2i}{(4\pi)^{2-\epsilon}}\frac{\Gamma(\epsilon+1)}{\Gamma(2)}\int^1_0dx \frac{F_2^{-\epsilon}}{R_1^{1-\epsilon}F_1} = 0,
\end{equation}
which evaluates to zero if we perform the integration over $x$. The term $I_{2}$ is zero for the same reason, we can compute it following the same steps. This means
\begin{equation}
    I^{\mathrm{virtual}}_{12} = 0.
\end{equation}

%%%%%%%%%%%%%%%%%%%%%%%%%%%%%%%%%%%%%%%%%%%%%%%%%%%%%%%%%%%%%%%%%%%%%%%%%%%%%%%%%%
\bibliographystyle{JHEP}
\normalbaselines 
\bibliography{TMD_ref}

\providecommand{\href}[2]{#2}\begingroup\raggedright\begin{thebibliography}{10}

\bibitem{Gao:2005iu}
Y.~Gao, C.~S. Li and J.~J. Liu, \emph{{Transverse momentum resummation for
  Higgs production in soft-collinear effective theory}},
  \href{http://dx.doi.org/10.1103/PhysRevD.72.114020}{\emph{Phys. Rev. D} {\bf
  72} (2005) 114020}, [\href{http://arxiv.org/abs/hep-ph/0501229}{{\tt
  hep-ph/0501229}}].

\bibitem{Chiu:2012ir}
J.-Y. Chiu, A.~Jain, D.~Neill and I.~Z. Rothstein, \emph{{A Formalism for the
  Systematic Treatment of Rapidity Logarithms in Quantum Field Theory}},
  \href{http://dx.doi.org/10.1007/JHEP05(2012)084}{\emph{JHEP} {\bf 05} (2012)
  084}, [\href{http://arxiv.org/abs/1202.0814}{{\tt 1202.0814}}].

\bibitem{Echevarria:2015uaa}
M.~G. Echevarria, T.~Kasemets, P.~J. Mulders and C.~Pisano, \emph{{QCD
  evolution of (un)polarized gluon TMDPDFs and the Higgs $q_T$-distribution}},
  \href{http://dx.doi.org/10.1007/JHEP07(2015)158}{\emph{JHEP} {\bf 07} (2015)
  158}, [\href{http://arxiv.org/abs/1502.05354}{{\tt 1502.05354}}].

\bibitem{Neill:2015roa}
D.~Neill, I.~Z. Rothstein and V.~Vaidya, \emph{{The Higgs Transverse Momentum
  Distribution at NNLL and its Theoretical Errors}},
  \href{http://dx.doi.org/10.1007/JHEP12(2015)097}{\emph{JHEP} {\bf 12} (2015)
  097}, [\href{http://arxiv.org/abs/1503.00005}{{\tt 1503.00005}}].

\bibitem{Gutierrez-Reyes:2019rug}
D.~Gutierrez-Reyes, S.~Leal-Gomez, I.~Scimemi and A.~Vladimirov,
  \emph{{Linearly polarized gluons at next-to-next-to leading order and the
  Higgs transverse momentum distribution}},
  \href{http://dx.doi.org/10.1007/JHEP11(2019)121}{\emph{JHEP} {\bf 11} (2019)
  121}, [\href{http://arxiv.org/abs/1907.03780}{{\tt 1907.03780}}].

\bibitem{Monni:2019yyr}
P.~F. Monni, L.~Rottoli and P.~Torrielli, \emph{{Higgs transverse momentum with
  a jet veto: a double-differential resummation}},
  \href{http://dx.doi.org/10.1103/PhysRevLett.124.252001}{\emph{Phys. Rev.
  Lett.} {\bf 124} (2020) 252001}, [\href{http://arxiv.org/abs/1909.04704}{{\tt
  1909.04704}}].

\bibitem{Chen:2018pzu}
X.~Chen, T.~Gehrmann, E.~N. Glover, A.~Huss, Y.~Li, D.~Neill et~al.,
  \emph{{Precise QCD Description of the Higgs Boson Transverse Momentum
  Spectrum}},
  \href{http://dx.doi.org/10.1016/j.physletb.2018.11.037}{\emph{Phys. Lett. B}
  {\bf 788} (2019) 425--430}, [\href{http://arxiv.org/abs/1805.00736}{{\tt
  1805.00736}}].

\bibitem{Mulders:2000sh}
P.~Mulders and J.~Rodrigues, \emph{{Transverse momentum dependence in gluon
  distribution and fragmentation functions}},
  \href{http://dx.doi.org/10.1103/PhysRevD.63.094021}{\emph{Phys. Rev. D} {\bf
  63} (2001) 094021}, [\href{http://arxiv.org/abs/hep-ph/0009343}{{\tt
  hep-ph/0009343}}].

\bibitem{Boer:2012bt}
D.~Boer and C.~Pisano, \emph{{Polarized gluon studies with charmonium and
  bottomonium at LHCb and AFTER}},
  \href{http://dx.doi.org/10.1103/PhysRevD.86.094007}{\emph{Phys. Rev. D} {\bf
  86} (2012) 094007}, [\href{http://arxiv.org/abs/1208.3642}{{\tt 1208.3642}}].

\bibitem{Ma:2012hh}
J.~Ma, J.~Wang and S.~Zhao, \emph{{Transverse momentum dependent factorization
  for quarkonium production at low transverse momentum}},
  \href{http://dx.doi.org/10.1103/PhysRevD.88.014027}{\emph{Phys. Rev. D} {\bf
  88} (2013) 014027}, [\href{http://arxiv.org/abs/1211.7144}{{\tt 1211.7144}}].

\bibitem{Zhang:2014vmh}
G.-P. Zhang, \emph{{Probing transverse momentum dependent gluon distribution
  functions from hadronic quarkonium pair production}},
  \href{http://dx.doi.org/10.1103/PhysRevD.90.094011}{\emph{Phys. Rev. D} {\bf
  90} (2014) 094011}, [\href{http://arxiv.org/abs/1406.5476}{{\tt 1406.5476}}].

\bibitem{Ma:2015vpt}
J.~Ma and C.~Wang, \emph{{QCD factorization for quarkonium production in hadron
  collisions at low transverse momentum}},
  \href{http://dx.doi.org/10.1103/PhysRevD.93.014025}{\emph{Phys. Rev. D} {\bf
  93} (2016) 014025}, [\href{http://arxiv.org/abs/1509.04421}{{\tt
  1509.04421}}].

\bibitem{Boer:2015uqa}
D.~Boer, \emph{{Linearly polarized gluon effects in unpolarized collisions}},
  \href{http://dx.doi.org/10.22323/1.249.0023}{\emph{PoS} {\bf QCDEV2015}
  (2015) 023}, [\href{http://arxiv.org/abs/1510.05915}{{\tt 1510.05915}}].

\bibitem{Bain:2016rrv}
R.~Bain, Y.~Makris and T.~Mehen, \emph{{Transverse Momentum Dependent
  Fragmenting Jet Functions with Applications to Quarkonium Production}},
  \href{http://dx.doi.org/10.1007/JHEP11(2016)144}{\emph{JHEP} {\bf 11} (2016)
  144}, [\href{http://arxiv.org/abs/1610.06508}{{\tt 1610.06508}}].

\bibitem{Mukherjee:2015smo}
A.~Mukherjee and S.~Rajesh, \emph{{Probing Transverse Momentum Dependent Parton
  Distributions in Charmonium and Bottomonium Production}},
  \href{http://dx.doi.org/10.1103/PhysRevD.93.054018}{\emph{Phys. Rev. D} {\bf
  93} (2016) 054018}, [\href{http://arxiv.org/abs/1511.04319}{{\tt
  1511.04319}}].

\bibitem{Mukherjee:2016cjw}
A.~Mukherjee and S.~Rajesh, \emph{{Linearly polarized gluons in charmonium and
  bottomonium production in color octet model}},
  \href{http://dx.doi.org/10.1103/PhysRevD.95.034039}{\emph{Phys. Rev. D} {\bf
  95} (2017) 034039}, [\href{http://arxiv.org/abs/1611.05974}{{\tt
  1611.05974}}].

\bibitem{Lansberg:2017tlc}
J.-P. Lansberg, C.~Pisano and M.~Schlegel, \emph{{Associated production of a
  dilepton and a $\Upsilon(J/\psi)$ at the LHC as a probe of gluon transverse
  momentum dependent distributions}},
  \href{http://dx.doi.org/10.1016/j.nuclphysb.2017.04.011}{\emph{Nucl. Phys. B}
  {\bf 920} (2017) 192--210}, [\href{http://arxiv.org/abs/1702.00305}{{\tt
  1702.00305}}].

\bibitem{Lansberg:2017dzg}
J.-P. Lansberg, C.~Pisano, F.~Scarpa and M.~Schlegel, \emph{{Pinning down the
  linearly-polarised gluons inside unpolarised protons using quarkonium-pair
  production at the LHC}},
  \href{http://dx.doi.org/10.1016/j.physletb.2018.08.004}{\emph{Phys. Lett. B}
  {\bf 784} (2018) 217--222}, [\href{http://arxiv.org/abs/1710.01684}{{\tt
  1710.01684}}].

\bibitem{Bacchetta:2018ivt}
A.~Bacchetta, D.~Boer, C.~Pisano and P.~Taels, \emph{{Gluon TMDs and NRQCD
  matrix elements in $J/\psi$ production at an EIC}},
  \href{http://dx.doi.org/10.1140/epjc/s10052-020-7620-8}{\emph{Eur. Phys. J.
  C} {\bf 80} (2020) 72}, [\href{http://arxiv.org/abs/1809.02056}{{\tt
  1809.02056}}].

\bibitem{Hadjidakis:2018ifr}
C.~Hadjidakis et~al., \emph{{A Fixed-Target Programme at the LHC: Physics Case
  and Projected Performances for Heavy-Ion, Hadron, Spin and Astroparticle
  Studies}},  \href{http://arxiv.org/abs/1807.00603}{{\tt 1807.00603}}.

\bibitem{DAlesio:2019qpk}
U.~D'Alesio, F.~Murgia, C.~Pisano and P.~Taels, \emph{{Azimuthal asymmetries in
  semi-inclusive $J/\psi\,+\,\mathrm{jet}$ production at an EIC}},
  \href{http://dx.doi.org/10.1103/PhysRevD.100.094016}{\emph{Phys. Rev. D} {\bf
  100} (2019) 094016}, [\href{http://arxiv.org/abs/1908.00446}{{\tt
  1908.00446}}].

\bibitem{Echevarria:2019ynx}
M.~G. Echevarria, \emph{{Proper TMD factorization for quarkonia production:
  $pp\to\eta_{c,b}$ as a study case}},
  \href{http://dx.doi.org/10.1007/JHEP10(2019)144}{\emph{JHEP} {\bf 10} (2019)
  144}, [\href{http://arxiv.org/abs/1907.06494}{{\tt 1907.06494}}].

\bibitem{Fleming:2019pzj}
S.~Fleming, Y.~Makris and T.~Mehen, \emph{{An effective field theory approach
  to quarkonium at small transverse momentum}},
  \href{http://dx.doi.org/10.1007/JHEP04(2020)122}{\emph{JHEP} {\bf 04} (2020)
  122}, [\href{http://arxiv.org/abs/1910.03586}{{\tt 1910.03586}}].

\bibitem{Scarpa:2019fol}
F.~Scarpa, D.~Boer, M.~G. Echevarria, J.-P. Lansberg, C.~Pisano and
  M.~Schlegel, \emph{{Studies of gluon TMDs and their evolution using
  quarkonium-pair production at the LHC}},
  \href{http://dx.doi.org/10.1140/epjc/s10052-020-7619-1}{\emph{Eur. Phys. J.
  C} {\bf 80} (2020) 87}, [\href{http://arxiv.org/abs/1909.05769}{{\tt
  1909.05769}}].

\bibitem{Grewal:2020hoc}
M.~Grewal, Z.-B. Kang, J.-W. Qiu and A.~Signori, \emph{{Predictive power of
  transverse-momentum-dependent distributions}},
  \href{http://dx.doi.org/10.1103/PhysRevD.101.114023}{\emph{Phys. Rev. D} {\bf
  101} (2020) 114023}, [\href{http://arxiv.org/abs/2003.07453}{{\tt
  2003.07453}}].

\bibitem{Boer:2020bbd}
D.~Boer, U.~D'Alesio, F.~Murgia, C.~Pisano and P.~Taels, \emph{{$J/\psi$ meson
  production in SIDIS: matching high and low transverse momentum}},
  \href{http://arxiv.org/abs/2004.06740}{{\tt 2004.06740}}.

\bibitem{Echevarria:2020qjk}
M.~G. Echevarria, Y.~Makris and I.~Scimemi, \emph{{Quarkonium TMD fragmentation
  functions in NRQCD}},  \href{http://arxiv.org/abs/2007.05547}{{\tt
  2007.05547}}.

\bibitem{Dominguez:2010xd}
F.~Dominguez, B.-W. Xiao and F.~Yuan, \emph{{$k_t$-factorization for Hard
  Processes in Nuclei}},
  \href{http://dx.doi.org/10.1103/PhysRevLett.106.022301}{\emph{Phys. Rev.
  Lett.} {\bf 106} (2011) 022301}, [\href{http://arxiv.org/abs/1009.2141}{{\tt
  1009.2141}}].

\bibitem{Zhu:2013yxa}
R.~Zhu, P.~Sun and F.~Yuan, \emph{{Low Transverse Momentum Heavy Quark Pair
  Production to Probe Gluon Tomography}},
  \href{http://dx.doi.org/10.1016/j.physletb.2013.11.002}{\emph{Phys. Lett. B}
  {\bf 727} (2013) 474--479}, [\href{http://arxiv.org/abs/1309.0780}{{\tt
  1309.0780}}].

\bibitem{Zhang:2017uiz}
G.-P. Zhang, \emph{{Back-to-back heavy quark pair production in Semi-inclusive
  DIS}}, \href{http://dx.doi.org/10.1007/JHEP11(2017)069}{\emph{JHEP} {\bf 11}
  (2017) 069}, [\href{http://arxiv.org/abs/1709.08970}{{\tt 1709.08970}}].

\bibitem{Boer:2010zf}
D.~Boer, S.~J. Brodsky, P.~J. Mulders and C.~Pisano, \emph{{Direct Probes of
  Linearly Polarized Gluons inside Unpolarized Hadrons}},
  \href{http://dx.doi.org/10.1103/PhysRevLett.106.132001}{\emph{Phys. Rev.
  Lett.} {\bf 106} (2011) 132001}, [\href{http://arxiv.org/abs/1011.4225}{{\tt
  1011.4225}}].

\bibitem{Chu:2017mnm}
X.~Chu, E.-C. Aschenauer, J.-H. Lee and L.~Zheng, \emph{{Photon structure
  studied at an Electron Ion Collider}},
  \href{http://dx.doi.org/10.1103/PhysRevD.96.074035}{\emph{Phys. Rev. D} {\bf
  96} (2017) 074035}, [\href{http://arxiv.org/abs/1705.08831}{{\tt
  1705.08831}}].

\bibitem{Dumitru:2018kuw}
A.~Dumitru, V.~Skokov and T.~Ullrich, \emph{{Measuring the Weizsäcker-Williams
  distribution of linearly polarized gluons at an electron-ion collider through
  dijet azimuthal asymmetries}},
  \href{http://dx.doi.org/10.1103/PhysRevC.99.015204}{\emph{Phys. Rev. C} {\bf
  99} (2019) 015204}, [\href{http://arxiv.org/abs/1809.02615}{{\tt
  1809.02615}}].

\bibitem{Zheng:2018ssm}
L.~Zheng, E.~Aschenauer, J.~Lee, B.-W. Xiao and Z.-B. Yin, \emph{{Accessing the
  gluon Sivers function at a future electron-ion collider}},
  \href{http://dx.doi.org/10.1103/PhysRevD.98.034011}{\emph{Phys. Rev. D} {\bf
  98} (2018) 034011}, [\href{http://arxiv.org/abs/1805.05290}{{\tt
  1805.05290}}].

\bibitem{Page:2019gbf}
B.~Page, X.~Chu and E.~Aschenauer, \emph{{Experimental Aspects of Jet Physics
  at a Future EIC}},
  \href{http://dx.doi.org/10.1103/PhysRevD.101.072003}{\emph{Phys. Rev. D} {\bf
  101} (2020) 072003}, [\href{http://arxiv.org/abs/1911.00657}{{\tt
  1911.00657}}].

\bibitem{Arratia:2020azl}
M.~Arratia, Y.~Furletova, T.~Hobbs, F.~Olness and S.~J. Sekula, \emph{{Charm
  jets as a probe for strangeness at the future Electron-Ion Collider}},
  \href{http://arxiv.org/abs/2006.12520}{{\tt 2006.12520}}.

\bibitem{Chudakov:2016ytj}
E.~Chudakov, D.~Higinbotham, C.~Hyde, S.~Furletov, Y.~Furletova, D.~Nguyen
  et~al., \emph{{Heavy quark production at an Electron-Ion Collider}},
  \href{http://dx.doi.org/10.1088/1742-6596/770/1/012042}{\emph{J. Phys. Conf.
  Ser.} {\bf 770} (2016) 012042}, [\href{http://arxiv.org/abs/1610.08536}{{\tt
  1610.08536}}].

\bibitem{Li:2020zbk}
H.~T. Li, Z.~L. Liu and I.~Vitev, \emph{{Heavy meson tomography of cold nuclear
  matter at the electron-ion collider}},
  \href{http://arxiv.org/abs/2007.10994}{{\tt 2007.10994}}.

\bibitem{Fickinger:2016rfd}
M.~Fickinger, S.~Fleming, C.~Kim and E.~Mereghetti, \emph{{Effective field
  theory approach to heavy quark fragmentation}},
  \href{http://dx.doi.org/10.1007/JHEP11(2016)095}{\emph{JHEP} {\bf 11} (2016)
  095}, [\href{http://arxiv.org/abs/1606.07737}{{\tt 1606.07737}}].

\bibitem{Anderle:2017cgl}
D.~P. Anderle, T.~Kaufmann, M.~Stratmann, F.~Ringer and I.~Vitev, \emph{{Using
  hadron-in-jet data in a global analysis of $D^{*}$ fragmentation functions}},
  \href{http://dx.doi.org/10.1103/PhysRevD.96.034028}{\emph{Phys. Rev. D} {\bf
  96} (2017) 034028}, [\href{http://arxiv.org/abs/1706.09857}{{\tt
  1706.09857}}].

\bibitem{Buffing:2018ggv}
M.~G. Buffing, Z.-B. Kang, K.~Lee and X.~Liu, \emph{{A transverse momentum
  dependent framework for back-to-back photon+jet production}},
  \href{http://arxiv.org/abs/1812.07549}{{\tt 1812.07549}}.

\bibitem{Chien:2019gyf}
Y.-T. Chien, D.~Y. Shao and B.~Wu, \emph{{Resummation of Boson-Jet Correlation
  at Hadron Colliders}},
  \href{http://dx.doi.org/10.1007/JHEP11(2019)025}{\emph{JHEP} {\bf 11} (2019)
  025}, [\href{http://arxiv.org/abs/1905.01335}{{\tt 1905.01335}}].

\bibitem{Echevarria:2015byo}
M.~G. Echevarria, I.~Scimemi and A.~Vladimirov, \emph{{Universal transverse
  momentum dependent soft function at NNLO}},
  \href{http://dx.doi.org/10.1103/PhysRevD.93.054004}{\emph{Phys. Rev.} {\bf
  D93} (2016) 054004}, [\href{http://arxiv.org/abs/1511.05590}{{\tt
  1511.05590}}].

\bibitem{Echevarria:2016scs}
M.~G. Echevarria, I.~Scimemi and A.~Vladimirov, \emph{{Unpolarized Transverse
  Momentum Dependent Parton Distribution and Fragmentation Functions at
  next-to-next-to-leading order}},
  \href{http://dx.doi.org/10.1007/JHEP09(2016)004}{\emph{JHEP} {\bf 09} (2016)
  004}, [\href{http://arxiv.org/abs/1604.07869}{{\tt 1604.07869}}].

\bibitem{Gutierrez-Reyes:2018qez}
D.~Gutierrez-Reyes, I.~Scimemi, W.~J. Waalewijn and L.~Zoppi, \emph{{Transverse
  momentum dependent distributions with jets}},
  \href{http://dx.doi.org/10.1103/PhysRevLett.121.162001}{\emph{Phys. Rev.
  Lett.} {\bf 121} (2018) 162001}, [\href{http://arxiv.org/abs/1807.07573}{{\tt
  1807.07573}}].

\bibitem{Gutierrez-Reyes:2019vbx}
D.~Gutierrez-Reyes, I.~Scimemi, W.~J. Waalewijn and L.~Zoppi, \emph{{Transverse
  momentum dependent distributions in $e^+e^-$ and semi-inclusive
  deep-inelastic scattering using jets}},
  \href{http://dx.doi.org/10.1007/JHEP10(2019)031}{\emph{JHEP} {\bf 10} (2019)
  031}, [\href{http://arxiv.org/abs/1904.04259}{{\tt 1904.04259}}].

\bibitem{Gutierrez-Reyes:2019msa}
D.~Gutierrez-Reyes, Y.~Makris, V.~Vaidya, I.~Scimemi and L.~Zoppi,
  \emph{{Probing Transverse-Momentum Distributions With Groomed Jets}},
  \href{http://dx.doi.org/10.1007/JHEP08(2019)161}{\emph{JHEP} {\bf 08} (2019)
  161}, [\href{http://arxiv.org/abs/1907.05896}{{\tt 1907.05896}}].

\bibitem{Arratia:2020ssx}
M.~Arratia, Y.~Makris, D.~Neill, F.~Ringer and N.~Sato, \emph{{Asymmetric jet
  clustering in deep-inelastic scattering}},
  \href{http://arxiv.org/abs/2006.10751}{{\tt 2006.10751}}.

\bibitem{Hornig:2016ahz}
A.~Hornig, Y.~Makris and T.~Mehen, \emph{{Jet Shapes in Dijet Events at the LHC
  in SCET}}, \href{http://dx.doi.org/10.1007/JHEP04(2016)097}{\emph{JHEP} {\bf
  04} (2016) 097}, [\href{http://arxiv.org/abs/1601.01319}{{\tt 1601.01319}}].

\bibitem{Luo:2019bmw}
M.-X. Luo, T.-Z. Yang, H.~X. Zhu and Y.~J. Zhu, \emph{{Transverse Parton
  Distribution and Fragmentation Functions at NNLO: the Gluon Case}},
  \href{http://dx.doi.org/10.1007/JHEP01(2020)040}{\emph{JHEP} {\bf 01} (2020)
  040}, [\href{http://arxiv.org/abs/1909.13820}{{\tt 1909.13820}}].

\bibitem{Echevarria:2012pw}
M.~G. Echevarria, A.~Idilbi, A.~Schaefer and I.~Scimemi,
  \emph{{Model-Independent Evolution of Transverse Momentum Dependent
  Distribution Functions (TMDs) at NNLL}},
  \href{http://dx.doi.org/10.1140/epjc/s10052-013-2636-y}{\emph{Eur. Phys. J.
  C} {\bf 73} (2013) 2636}, [\href{http://arxiv.org/abs/1208.1281}{{\tt
  1208.1281}}].

\bibitem{Echevarria:2014rua}
M.~G. Echevarria, A.~Idilbi and I.~Scimemi, \emph{{Unified treatment of the QCD
  evolution of all (un-)polarized transverse momentum dependent functions:
  Collins function as a study case}},
  \href{http://dx.doi.org/10.1103/PhysRevD.90.014003}{\emph{Phys. Rev.} {\bf
  D90} (2014) 014003}, [\href{http://arxiv.org/abs/1402.0869}{{\tt
  1402.0869}}].

\bibitem{Li:2016ctv}
Y.~Li and H.~X. Zhu, \emph{{Bootstrapping Rapidity Anomalous Dimensions for
  Transverse-Momentum Resummation}},
  \href{http://dx.doi.org/10.1103/PhysRevLett.118.022004}{\emph{Phys. Rev.
  Lett.} {\bf 118} (2017) 022004}, [\href{http://arxiv.org/abs/1604.01404}{{\tt
  1604.01404}}].

\bibitem{Vladimirov:2016dll}
A.~A. Vladimirov, \emph{{Soft-/rapidity- anomalous dimensions correspondence}},
  \href{http://dx.doi.org/10.1103/PhysRevLett.118.062001}{\emph{Phys. Rev.
  Lett.} {\bf 118} (2017) 062001}, [\href{http://arxiv.org/abs/1610.05791}{{\tt
  1610.05791}}].

\bibitem{Scimemi:2019cmh}
I.~Scimemi and A.~Vladimirov, \emph{{Non-perturbative structure of
  semi-inclusive deep-inelastic and Drell-Yan scattering at small transverse
  momentum}}, \href{http://dx.doi.org/10.1007/JHEP06(2020)137}{\emph{JHEP} {\bf
  06} (2020) 137}, [\href{http://arxiv.org/abs/1912.06532}{{\tt 1912.06532}}].

\bibitem{Becher:2009th}
T.~Becher and M.~D. Schwartz, \emph{{Direct photon production with effective
  field theory}}, \href{http://dx.doi.org/10.1007/JHEP02(2010)040}{\emph{JHEP}
  {\bf 02} (2010) 040}, [\href{http://arxiv.org/abs/0911.0681}{{\tt
  0911.0681}}].

\bibitem{Becher:2012xr}
T.~Becher, C.~Lorentzen and M.~D. Schwartz, \emph{{Precision Direct Photon and
  W-Boson Spectra at High p\_T and Comparison to LHC Data}},
  \href{http://dx.doi.org/10.1103/PhysRevD.86.054026}{\emph{Phys. Rev. D} {\bf
  86} (2012) 054026}, [\href{http://arxiv.org/abs/1206.6115}{{\tt 1206.6115}}].

\bibitem{Chien:2020hzh}
Y.-T. Chien, R.~Rahn, S.~Schrijnder~van Velzen, D.~Y. Shao, W.~J. Waalewijn and
  B.~Wu, \emph{{Azimuthal angle for boson-jet production in the back-to-back
  limit}},  \href{http://arxiv.org/abs/2005.12279}{{\tt 2005.12279}}.

\bibitem{GarciaEchevarria:2011rb}
M.~G. Echevarria, A.~Idilbi and I.~Scimemi, \emph{{Factorization Theorem For
  Drell-Yan At Low $q_T$ And Transverse Momentum Distributions
  On-The-Light-Cone}},
  \href{http://dx.doi.org/10.1007/JHEP07(2012)002}{\emph{JHEP} {\bf 07} (2012)
  002}, [\href{http://arxiv.org/abs/1111.4996}{{\tt 1111.4996}}].

\bibitem{Collins:2011zzd}
J.~Collins, \emph{{Foundations of perturbative QCD}}.
\newblock Cambridge University Press, 2013.

\bibitem{Echevarria:2012js}
M.~G. Echevarria, A.~Idilbi and I.~Scimemi, \emph{{Soft and Collinear
  Factorization and Transverse Momentum Dependent Parton Distribution
  Functions}},
  \href{http://dx.doi.org/10.1016/j.physletb.2013.09.003}{\emph{Phys. Lett.}
  {\bf B726} (2013) 795--801}, [\href{http://arxiv.org/abs/1211.1947}{{\tt
  1211.1947}}].

\bibitem{Jaffe:1993ie}
R.~Jaffe and L.~Randall, \emph{{Heavy quark fragmentation into heavy mesons}},
  \href{http://dx.doi.org/10.1016/0550-3213(94)90495-2}{\emph{Nucl. Phys. B}
  {\bf 412} (1994) 79--105}, [\href{http://arxiv.org/abs/hep-ph/9306201}{{\tt
  hep-ph/9306201}}].

\bibitem{Fleming:2007qr}
S.~Fleming, A.~H. Hoang, S.~Mantry and I.~W. Stewart, \emph{{Jets from massive
  unstable particles: Top-mass determination}},
  \href{http://dx.doi.org/10.1103/PhysRevD.77.074010}{\emph{Phys. Rev. D} {\bf
  77} (2008) 074010}, [\href{http://arxiv.org/abs/hep-ph/0703207}{{\tt
  hep-ph/0703207}}].

\bibitem{Fleming:2007xt}
S.~Fleming, A.~H. Hoang, S.~Mantry and I.~W. Stewart, \emph{{Top Jets in the
  Peak Region: Factorization Analysis with NLL Resummation}},
  \href{http://dx.doi.org/10.1103/PhysRevD.77.114003}{\emph{Phys. Rev. D} {\bf
  77} (2008) 114003}, [\href{http://arxiv.org/abs/0711.2079}{{\tt 0711.2079}}].

\bibitem{Boglione:2020cwn}
M.~Boglione and A.~Simonelli, \emph{{Universality-breaking effects in $e^+e^-$
  hadronic production processes}},  \href{http://arxiv.org/abs/2007.13674}{{\tt
  2007.13674}}.

\bibitem{Kang:2020xez}
Z.-B. Kang, K.~Lee, D.~Y. Shao and J.~Terry, \emph{{The Sivers Asymmetry in
  Hadronic Dijet Production}},  \href{http://arxiv.org/abs/2008.05470}{{\tt
  2008.05470}}.

\bibitem{Ellis:2010rwa}
S.~D. Ellis, C.~K. Vermilion, J.~R. Walsh, A.~Hornig and C.~Lee, \emph{{Jet
  Shapes and Jet Algorithms in SCET}},
  \href{http://dx.doi.org/10.1007/JHEP11(2010)101}{\emph{JHEP} {\bf 11} (2010)
  101}, [\href{http://arxiv.org/abs/1001.0014}{{\tt 1001.0014}}].

\end{thebibliography}\endgroup
%%%%%%%%%%%%%%%%%%%%%%%%%%%%%%%%%%%%%%%%%%%%%%%%%%%%%%%%%%%%%%%%%%%%%%%%%%%%%%%%%%

\end{document}